\documentclass[aip,longbiblography
]{revtex4-2}
\usepackage{chemformula}
\usepackage{graphicx}
\usepackage{amsmath,amsfonts,amssymb}
\draft 
\usepackage{mathrsfs}
\usepackage[scr=rsfs,cal=boondox]{mathalfa}
\begin{document}


\title[Molecular Beam Epitaxy of \ch{KTaO3}]{Molecular Beam Epitaxy of \ch{KTaO3}} 



\author{Tobias Schwaigert}
\affiliation{Platform for the Accelerated Realization, Analysis, and Discovery of Interface Materials (PARADIM), Cornell University, Ithaca, New York 14853, USA}
\affiliation{Department of Materials Science and Engineering, Cornell University, Ithaca, NY 14853, USA}
\author{Salva Salmani-Rezaie}
\affiliation{School of Applied and Engineering Physics, Cornell University, Ithaca, New York 14853, USA}
\affiliation{Kavli Institute at Cornell for Nanoscale Science, Cornell University, Ithaca, New York 14853, USA}
\author{Matthew R. Barone}
\affiliation{Platform for the Accelerated Realization, Analysis, and Discovery of Interface Materials (PARADIM), Cornell University, Ithaca, New York 14853, USA}
\affiliation{Department of Materials Science and Engineering, Cornell University, Ithaca, NY 14853, USA}
\author{Hanjong Paik}
\affiliation{Platform for the Accelerated Realization, Analysis, and Discovery of Interface Materials (PARADIM), Cornell University, Ithaca, New York 14853, USA}
\affiliation{School of Electrical \& Computer Engineering, University of Oklahoma, Norman, OK 73019, USA}
\affiliation{Center for Quantum Research and Technology, University of Oklahoma, Norman, OK 73019, USA}
\author{Ethan Ray}
\affiliation{Platform for the Accelerated Realization, Analysis, and Discovery of Interface Materials (PARADIM), Cornell University, Ithaca, New York 14853, USA}
\author{Michael D. Williams}
\affiliation{Department of Physics, Clark Atlanta University, Atlanta, Georgia 30314, USA}
\author{David A. Muller}
\affiliation{School of Applied and Engineering Physics, Cornell University, Ithaca, New York 14853, USA}
\affiliation{Kavli Institute at Cornell for Nanoscale Science, Cornell University, Ithaca, New York 14853, USA}
\author{Darrell G. Schlom}
\affiliation{Platform for the Accelerated Realization, Analysis, and Discovery of Interface Materials (PARADIM), Cornell University, Ithaca, New York 14853, USA}
\affiliation{Department of Materials Science and Engineering, Cornell University, Ithaca, NY 14853, USA}
\affiliation{Kavli Institute at Cornell for Nanoscale Science, Cornell University, Ithaca, New York 14853, USA}
\affiliation{Leibniz-Institut für Kristallzüchtung, Max-Born-Str. 2, 12489 Berlin, Germany}
\author{Kaveh Ahadi}
\email[kahadi@ncsu.edu]{}
\affiliation{Department of Materials Science and Engineering, North Carolina State University, Raleigh, NC 27265 USA}
\affiliation{Department of Physics, North Carolina State University, Raleigh, North Carolina 27695, USA}

\date{\today}

\begin{abstract}
 Strain-engineering is a powerful means to tune the polar, structural, and electronic instabilities of incipient ferroelectrics. \ch{KTaO3} is near a polar instability and shows anisotropic superconductivity in electron-doped samples. Here, we demonstrate growth of high quality \ch{KTaO3} thin films by molecular-beam epitaxy. Tantalum was provided by both a suboxide source emanating a \ch{TaO2} flux from \ch{Ta2O5} contained in a conventional effusion cell as well as an electron-beam-heated tantalum source. Excess potassium and  a combination of ozone and oxygen (10 \% \ch{O3} + 90 \% \ch{O2}) were simultaneously supplied with the \ch{TaO2} (or tantalum) molecular beams to grow the \ch{KTaO3} films. Laue fringes suggest that the films are smooth with an abrupt film/substrate interface. Cross-sectional scanning transmission electron microscopy does not show any extended defects and confirms that the films have an atomically abrupt interface with the  substrate. Atomic force microscopy reveals atomic steps at the surface of the grown films. Reciprocal space mapping demonstrates that the films, when sufficiently thin, are coherently strained to the \ch{SrTiO3} (001) and \ch{GdScO3} (110) substrates.
\end{abstract}


\maketitle 

\section{Introduction\label{sec:level1}}

Complex transition metal oxides exhibit a broad spectrum of orders and instabilities. Tuning  the rich and often record properties of these materials is facilitated by their incorporation in high-quality epitaxial heterostructures where strain, juxtaposed competing orders, or other methodologies to modify the ground state can be imposed.\cite{haeni2004room,ahadi2019enhancing, lee2010strong, ahadi2017evidence} \ch{KTaO3} is an incipient ferroelectric, in which superconductivity emerges at low temperatures in electron-doped samples.\cite{ueno2011discovery,changjiang,gupta2022ktao3} The \ch{KTaO3} conduction band is derived from the Ta 5\emph{d} states and shows highly anisotropic electronic transport.\cite{al2021two, al2022oxygen, al2022superconductivity} Furthermore, charge carriers in \ch{KTaO3} have smaller effective mass and larger spin–orbit coupling compared to \ch{SrTiO3}.\cite{nakamura2009,wadehra2020} These opportunities invite the intensive study of \ch{KTaO3}-based thin films and heterostructures to understand and engineer these phenomena. Surprising, the growth of \ch{KTaO3} by molecular-beam epitaxy (MBE) has not been demonstrated. 

The main challenges to the  MBE growth of \ch{KTaO3} are to provide a stable tantalum flux and the high chemical reactivity between potassium metal and air that complicates the use of elemental potassium as an MBE source. Tantalum is a refractory metal, requiring temperatures in excess of 2600 °C to evaporate at typical oxide MBE growth rates.\cite{honig1969} Successful MBE growth of tantalates remains elusive and has been limited to the use of electron-beam (e-beam) evaporator sources to reach the temperatures needed to evaporate elemental tantalum. This approach has been used to grow \ch{LiTaO3}\cite{sitar1995} and more recently \ch{Ta2SnO6}.\cite{barone2022} Recent thermodynamic calculations, however, suggest \ch{Ta2O5} as a potential source for the MBE growth of tantalates that can be accomplished at temperatures attainable in an MBE effusion cell.\cite{adkison2020} Elemental potassium, is highly reactive and readily oxidizes in air. A means to circumvent this issue is through the preparation of intermetallic compounds of alkali metals in a glove box with relatively low vapor pressure elements, e.g., \ch{LiSn4} and \ch{CsIn3}, as have been recently explored as  MBE sources.\cite{du2020control,parzyck2022}

 Here, we demonstrate the MBE growth of high-quality \ch{KTaO3} films  using both traditional elemental tantalum in an e-beam evaporator as well as a \ch{Ta2O5} source contained in a high-temperature MBE effusion cell. Potassium was evaporated from an \ch{In4K} intermetallic compound source. A combination of ozone and oxygen (10 \% \ch{O3} + 90 \% \ch{O2}) was used as the oxidant. The structural quality of the epitaxial \ch{KTaO3} films grown in multiple strain states was assessed using a wide range of characterization techniques. Although we have grown roughly fifty \ch{KTaO3} films with comparable quality, only the best three samples are featured in this article.
\section{Experimental}
Epitaxial \ch{KTaO3} films were grown in a Vecco GEN 10 MBE system. A molecular beam of \ch{TaO2} (gas) flux was generated from an effusion cell containing \ch{Ta2O5} (Alfa Aesar, 99.993 \%) contained in an iridium crucible. \ch{TaO2} is the most volatile species in the growth temperature range.\cite{adkison2020} Potassium was evaporated from an effusion cell, containing intermetallic \ch{In4K}, which melts at elevated temperature compared to pure potassium (432 °C vs 63.5 °C), improving the temperature control and flux stability.\cite{okamoto1992k}  The K-In alloy was prepared in a glove box and contained in a titanium crucible. Once prepared, it can be exposed to air, facilitating its handling and loading. The vapor pressure of potassium is more than $10^{10}$ times higher than indium at the K-In cell temperature of 300-400 °C.\cite{honig1969} \ch{GdScO3} (110)$_o$ (Crystec GmbH) substrates were used as received and the  \ch{SrTiO3} (001) substrates were terminated following the procedure developed by Koster \textit{et al.}\cite{koster1998} Films were grown by co-deposition of potassium, tantalum, and ozone at a substrate  temperature of 625 °C as measured by an optical pyrometer operating at a wavelenght of 1550 nm. The pyrometer measures the temperature of the platinum coating that has been evaporated on the backside of the substrate to facilitate radiative heat transfer from the SiC heating element of the MBE system to the substrate. The K:Ta flux ratio was kept at approximately 10:1. A mixture of  ozone and oxygen  (10 \% \ch{O3} + 90 \% \ch{O2}) was used as the oxidant. The films were grown at an oxidant background pressure of $1\times10^{-6}$ Torr. Typical fluxes for the the sources were (4-7)$\times 10^{12}$ atoms/$\text{cm}^2$/s for \ch{TaO2} and (4-7)$\times 10^{13}$ atoms/$\text{cm}^2$/s for potassium, determined by a quartz crystal microbalance (QCM), with an accuracy of about $\pm$ 15 \%. In a typical growth experiment the potassium flux was measured first, followed by \ch{TaO2} to ensure that the QCM was as close to RT as possible for the most accurate reading. For a more detailed description the reader is referred to the supplementary material. Codeposition with these fluxes results in a \ch{KTaO3} film growth rate of about 0.03 \AA/s.  

X-ray diffraction (XRD), X-ray reflectometry (XRR), and reciprocal space mapping (RSM) measurements were carried out using a PANalytical Empyrean diffractometer with Cu K$\alpha_{1}$ radiation. The raw XRR spectra were analyzed using the PANalytical X´Pert Reflectivity software package and the layer thickness was derived from a fast Fourier transform (FFT) after manually defining the critical angle to account for refractive effects. \textit{In situ} reflection high-energy electron diffraction (RHEED) patterns were recorded using KSA-400 software and a Staib electron source operated  at  14 kV and a filament current of 1.5 A. The morphology of the film surface was characterized using an Asylum Cypher ES environmental AFM. Cross-sectional scanning transmission electron microscopy (STEM) samples were prepared using standard lift-out process using a Thermo Fisher Scientific Helios G4UX focused ion beam with the final milling voltage of 2 kV for the gallium ions. A Thermo Fisher Scientific Spectra 300 X-CFEG, operating at 200 kV with a convergence angle of 30 mrad and a high-angle annular dark-field (HAADF) detector with an angular range of 60-200 mrad, was used to collect atomic resolution HAADF-STEM images. STEM energy-dispersive X-ray spectroscopy (EDX) data were collected using a steradian Dual-X EDX detector with a probe current of 100 pA. The noise of the STEM-EDX spectrum was reduced by the application of principal component analysis.
\section{Results}
\begin{figure*}[t]
\includegraphics[width=0.8\textwidth]{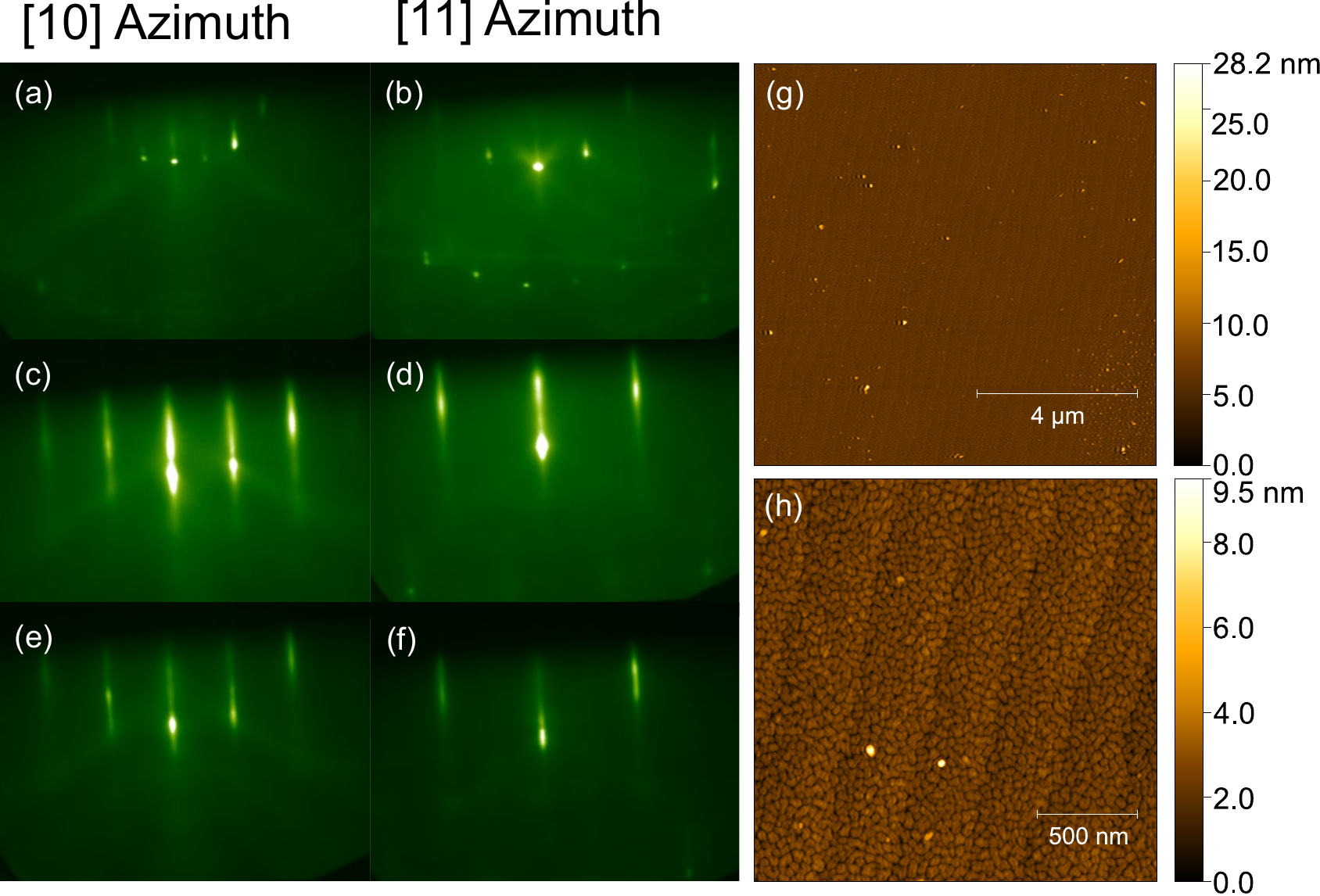}
\caption{\label{fig:RHEED} RHEED patterns of (a),(b) bare \ch{GdScO3} substrate; (c),(d) after 1 min (approximately 0.5 unit cell average thickness) \ch{KTaO3} growth; and (e),(f) immediately after the growth of an 18 nm thick \ch{KTaO3} film with an effusion cell containing \ch{Ta2O5}. (g),(h) Atomic force microscopy images at different magnifications, revealing atomic steps.}
\end{figure*}

\ch{KTaO3} is a cubic perovskite with a lattice constant  of $a_{KTO}$ = 3.988 \AA \cite{zhurova2000} at room temperature. The lattice mismatch between \ch{KTaO3} and \ch{GdScO3} (pseudo-cubic lattice-constant, 3.967 \AA \cite{uecker2008}) and \ch{SrTiO3} ($a_{STO}$ = 3.905 \AA \cite{culbertson2020}) are  –0.5\% and -2.1\%, respectively.
\ch{KTaO3} grows ``cube-on-pseudocube" on \ch{GdScO3} (110)$_o$ substrates. Reflection high-energy electron diffraction (RHEED) was used to monitor the evolution of the surface structure and reconstruction during growth. Figures~\ref{fig:RHEED}(a) and  ~\ref{fig:RHEED}(b) show the \ch{GdScO3} (110)$_o$ substrate along the high symmetry directions where diffraction streaks and Kikuchi lines are visible. Figures~\ref{fig:RHEED}(c) and~\ref{fig:RHEED}(d) show the diffraction pattern 1 min after the start of the growth (corresponding to the deposition of a \ch{KTaO3} film about one half of a unit cell thick on average) using the suboxide source. Kikuchi lines are still visible, but the diffraction pattern has evolved to \ch{KTaO3} (001). The RHEED pattern appears cloudy, suggesting a floating potassium oxide layer. The films are grown in a K-rich regime with a K:Ta flux ratio of 10:1 within an absorption-controlled growth regime exploiting the volatility of the potassium oxide species on the growth surface. Deviation from this flux ratio increases the roughness of the films. Figures~\ref{fig:RHEED}(e) and \ref{fig:RHEED}(f) show the \ch{KTaO3} RHEED streaks immediately after the growth of an 18 nm thick \ch{KTaO3} film, where the shutters of the \ch{TaO2} and potassium sources have been closed, but the substrate is still immersed in ozone and beginning to be cooled down from the growth temperature. Atomic force microscopy (AFM) images are shown in Figs.~\ref{fig:RHEED}(g) and  ~\ref{fig:RHEED}(h) at different magnifications. Atomic steps from the $<$ 0.05° off-cut substrate are visible. The root-mean-square (rms) roughness for Fig.~\ref{fig:RHEED}(h) is $\approx$ 640 pm, measured by taking a 1 $\mu\text{m}^2$ area as a reference. 

\begin{figure*}[t]
\includegraphics[width=\textwidth]{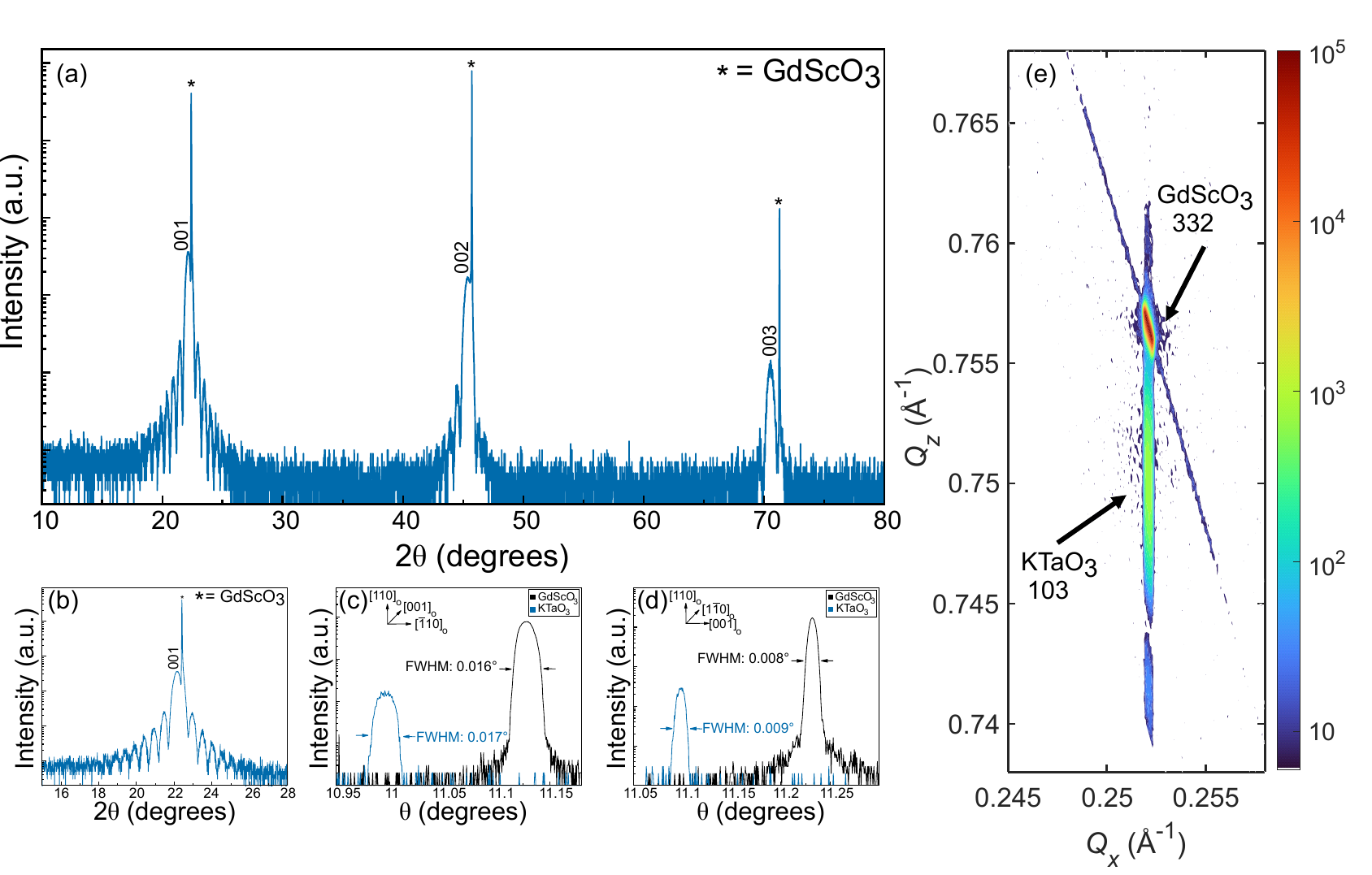}
\caption{\label{fig:xrd}X-ray diffraction of the 18 nm thick \ch{KTaO3} film grown on  a \ch{GdScO3} (110)$_o$ substrate with an effusion cell containing \ch{Ta2O5} . (a) $\theta$-2$\theta$ scan, showing 00$\mathcal{l}$ peaks of \ch{KTaO3}. Symmetric Laue fringes indicate a well-defined film thickness, indicative of an abrupt interface between film and substrate (asterisks * denote substrate reflections). (b) A zoomed-in $\theta$-2$\theta$ scan in the vicinity of  the \ch{KTaO3} 001 peak, showing the Laue fringes used to calculate the film thickness. (c),(d) Overlaid rocking curves of the  110 \ch{GdScO3} and 
 001 \ch{KTaO3} peaks, showing comparable FWHMs, indicating low out-of-plane mosaicity ($\Delta\omega\approx$ 0.017$^{\circ}$ and 0.008$^{\circ}$ along the two orthogonal in-plane directions of the substrate). (e) reciprocal space map (RSM) around the 332 substrate and 103 film reflections. The RSM results confirm that the film is fully strained to the substrate.} 
\end{figure*}

Figure~\ref{fig:xrd} shows the X-ray diffraction results of the same  18 nm thick \ch{KTaO3} film grown on a \ch{GdScO3} (110)$_o$ substrate using the suboxide source. The film thickness is calculated using the Laue fringes and corroborated with X-ray reflectivity and cross sectional HAADF-STEM. The $\theta$-2$\theta$ XRD scan only shows 00$\mathcal{l}$ peaks, confirming that the film is single-phase and oriented with its \textit{c}-axis perpendicular to the plane of the substrate. Figure~\ref{fig:xrd}(b) depicts a  close-up  $\theta$-2$\theta$ scan around the \ch{KTaO3} 001 peak, showing symmetric Laue fringes. The rocking curve full width at half maximum (FWHM) of the \ch{KTaO3} film is comparable to the \ch{GdScO3} substrate (both about 30 and 60 arc sec, respectively, along the two orthogonal in-plane directions of the substrate), suggesting the high crystalline quality of the grown films. X-ray reciprocal space mapping (RSM) around the \ch{GdScO3} 332 and \ch{KTaO3} 103 reflections confirms that the film is coherently strained to the substrate. 

 We apply elasticity theory to see how the observed out-of-plane lattice spacing of the commensurately strained \ch{KTaO3} films compares to a calculation using the elastic stiffness tensor of \ch{KTaO3}.\cite{LandoltBornstein1990} The out-of-plane lattice~ $a_\perp$ can be calculated from the out-of-plane strain, $\epsilon_{33}=\frac{(a_{\perp}-a_{\text{KTaO}})}{a_{\text{KTaO}}}$  by expanding the tensor equation (in Einstein notation): $\sigma_{ij} = c_{ijkl}\epsilon_{kl}$ for $\sigma_{33}$ and recognizing that $\sigma_{33}=0$ because the film is free of stress in the out-of-plane direction. This leads to: 
\begin{equation}
    a_\perp=a_{KTO}+\frac{2 (a_{KTO}-a_{STO})c_{12}}{c_{11}},
\end{equation}
where $c_{11}$ and $c_{12}$ are elastic stiffness tensor coefficients of \ch{KTaO3} in Voigt notation and $a_{KTO}$ and $a_{STO}$ are the lattice constants of unstrained \ch{KTaO3} and \ch{SrTiO3}, respectively. The calculated out-of-plane lattice constant expected for a commensurately strained \ch{KTaO3} film on \ch{SrTiO3} at room temperature is 4.028 \AA. This is  lower then the 4.043 $\pm$ 0.015\AA~ value measured by X-ray diffraction for the commensurately strained 10.5 nm thick \ch{KTaO3} film shown in Figs.~S1 and S2 (which was grown using an elemental tantalum molecular beam). 

\begin{figure}[t]
\includegraphics[width=0.5\textwidth]{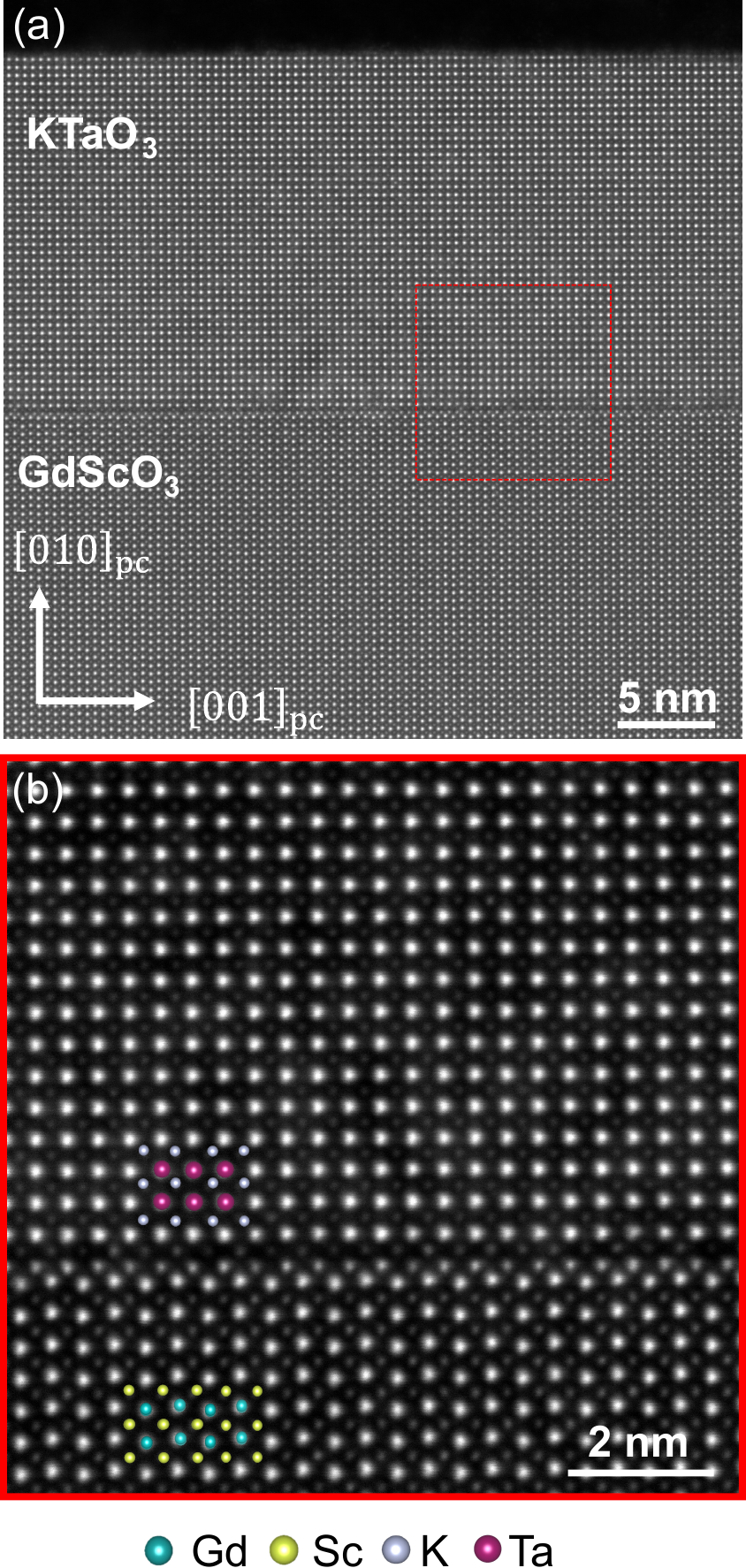}
\caption{\label{fig:STEM} (a) Cross-section HAADF-STEM images of the same 18 nm thick \ch{KTaO3} film grown on a \ch{GdScO3}(110) substrate with an effusion cell containing \ch{Ta2O5}. (b) Higher magnification HAADF-STEM image of the  \ch{KTaO3}/\ch{GdScO3} interface showing the bilayer of the intermixed metal ions.}
\end{figure}

In contrast to the extended out-of-plane lattice spacing observed for the commensurately strained \ch{KTaO3} film grown on a \ch{SrTiO3} substrate, the  18.0 nm thick commensurately strained \ch{KTaO3}/\ch{GdScO3} shown in Figs. \ref{fig:RHEED}-\ref{fig:STEM} shows the expected out-of-plane spacing, calculated with the elastic theory. Because \ch{GdScO3} is orthorhombic, the in-plane biaxial strains $\epsilon_{11}$ and $\epsilon_{22}$ imposed by the substrate are no longer equal and the equation for $a_\perp$ becomes,
 \begin{equation}
a_\perp =a_{KTO}+\frac{(4a_{KTO}-a_{GSO_{001}} -a_{GSO_{1\overline{1}0}})c_{12}}{2c_{11}},
 \end{equation}
where  $a_{GSO_{001}}$ and $a_{GSO_{1\overline{1}0}}$ are the in-plane distances that establish $\epsilon_{11}$ and $\epsilon_{22}$ through commensurate strain. Specifically, $a_{GSO_{001}}$ is the \textit{c}-axis length of \ch{GdScO3} (7.9314 \AA) and  $a_{GSO_{1\overline{1}0}}$ is the [$1\overline{1}0$] of \ch{GdScO3} (7.9401 \AA),\cite{uecker2013} where we are using the non-standard \textit{Pbnm} setting of \ch{GdScO3} as is most common in the literature. Here the calculations result in an expected spacing of  3.998 \AA~ at room temperature compared to the 3.997 $\pm$0.01 \AA~ measured by X-ray diffraction. The films grown in \ch{GdScO3} do not show any discrepancy between the measured and the calculated out-of-plane lattice parameter. Interestingly, \ch{KTaO3} films grown on \ch{SrTiO3} do show a discrepancy between the measured and calculated out-of-plane lattice parameter, which could be explained by the emergence of a ferroelectric state in films grown on \ch{SrTiO3}. Another possible explanation could be that the film grown on \ch{SrTiO3} might be non-stoichmetric. Errors in stoichiometry are known to lengthen the lattice constants of many perovskites, e.g., \ch{SrTiO3},\cite{brooks2009,jalan2009} \ch{CaTiO3},\cite{haislmaier2016} and \ch{SrVO3},\cite{moyer2013} but in other cases, e.g., \ch{LaVO3},\cite{zhang2015} can shorten them. The samples shown here are grown in an absorption controlled growth regime, yielding phase-pure \ch{KTaO3}. Nonetheless, phase purity is not synonymous with a stoichiometric \ch{KTaO3} film and it is possible that the growth conditions we have employed lead to non-stoichiometric \ch{KTaO3} films. If it is due to non-stoichiometry, the significant lattice expansion observed might be expected to give rise to extended defects as is the case for Sr-rich \ch{SrTiO3}  films.\cite{brooks2009} Low-magnification HAADF-STEM, however, does not show any extended defects in these films and can be found in the supplementary material in Fig. S7. A more interesting possibility is that the lattice expansion is intrinsic and is due to the \ch{KTaO3} under biaxial compression becoming ferroelectric with an out-of-plane polarization. For the \ch{KTaO3}/\ch{SrTiO3} system, first-principles calculations find that biaxial compressive strains of magnitude larger then 1 \% are needed to induce ferroelectricity.\cite{tyunina2010} This could elongate the out-of-plane lattice constant beyond that expected from an elasticity calculation because the ground state has changed from paraelectric \ch{KTaO3} to ferroelectric \ch{KTaO3} for the film commensurately strained to \ch{SrTiO3} (-2.1 \% strain), but not when \ch{KTaO3} is grown on \ch{GdScO3} (-0.5 \% strain). Future studies are planned to investigate this possibility.

Figure S1 compares the XRD of the \ch{KTaO3} films grown using a \ch{TaO2} suboxide molecular beam and a tantalum molecular beam from an e-beam-heated elemental tantalum  source.
Both \ch{KTaO3} films were grown on \ch{SrTiO3} (001) substrates at similar substrate temperature and ozone partial pressure.  Figures S2 and S3 show the X-ray diffraction $\theta$-2$\theta$ scans, RSM, and AFM characterization of these same films grown using suboxide and tantalum e-beam sources. Interestingly, the \ch{KTaO3} film, using the tantalum e-beam source, is strained to the \ch{SrTiO3} (001) substrate. This is noteworthy due to the large lattice mismatch between \ch{KTaO3} and \ch{SrTiO3} ($\approx$ -2.1 \%). The surface morphology revealed by  AFM, shows a smoother surface for the e-beam film. The AFM image of the suboxide film shows potassium oxide (\ch{K2O}) residues on the surface. This could be due to the presence of additional oxygen which oxidizes the potassium atoms, leading to a higher sticking coefficient. The rms roughness of the e-beam film is $\approx$ 0.3 nm compared to $\approx$ 1.8 nm for the suboxide film. The RSM around the \ch{SrTiO3} and \ch{KTaO3} 103 peak shows that only the e-beam sample is commensurately strained and the suboxide film is partially relaxed. The difference could be simply due to the difference in thickness:  22.5 nm and 10.5 nm for the suboxide and e-beam films, respectively. It is important to point out that an equivalent surface roughness can be achieved with the suboxide source and in thicker films (See Fig. \ref{fig:RHEED}).The initial results showed rougher surfaces with suboxide sources. After fine-tuning the growth parameters, we find that the suboxide sources also produce films that are atomically flat similar to those produced with the tantalum e-beam source. The flux emanating from the  \ch{Ta2O5} source is not only far more stable than the flux produced by the tantalum e-beam source. Furthermore, the suboxide flux can be increased to produce growth rates up to 100 nm/h. For these reasons we find the \ch{Ta2O5} source preferable for the growth of \ch{KTaO3} films by MBE.

High-angle annular dark-field scanning transmission electron microscope (HAADF-STEM) was used to further investigate the  \ch{KTaO3} films. The HAADF-STEM image along the \ch{GdScO3}[100]$_{pc}$ zone axis (Fig.~\ref{fig:STEM}), where the subscript \textit{pc} denotes pseudocubic indices, shows a coherent epitaxial interface between the \ch{KTaO3} film and underlying \ch{GdScO3} substrate. \ch{KTaO3} and \ch{GdScO3} both have polar surfaces and  the formation of two layers of  intermixed metal ions can relieve the polar catastrophe at the interface. Figure \ref{fig:STEM}(b) shows that the interface has the proposed\cite{thompson2014} bilayer structure with K$_{x}$Gd$_{1-x}$O(top)/Ta$_{y}$Sc$_{1-y}$O$_{2}$(bottom) to relieve the polarity conflict of the \ch{KTaO3} and \ch{GdScO3} interface. STEM-EDX analysis of the interface (Fig. S5) and the intensity line profile of the HAADF-STEM images (Fig. S6) also point to the formation of the intermixed bilayer structure. 
 
In summary, we demonstrate the  MBE growth of high-quality \ch{KTaO3} films. Suboxide and tantalum e-beam sources are used and compared. Potassium, evaporated from an \ch{In4K} compound source, provides reasonable flux stability. Symmetric Laue fringes suggest that the films are smooth. Cross section HAADF-STEM does not show any extended defects and reveals an atomically abrupt film/substrate interface. RSM confirms that when sufficiently thin the films are coherently strained to the substrates. The repeatability of the results and observed lattice spacings that are consistent with the stoichiometric growth of \ch{KTaO3} for strains where ferroelectricity is not expected, i.e., \ch{KTaO3}/\ch{GdScO3} (110), suggest that the growth by codeposition occurs in an absorption-controlled regime.

See supplementary material at [URL will be inserted by AIP Publishing] for Figs. S1–S9, a description of
the flux calibration and \ch{KTaO3} film growth and characterization.

 \begin{acknowledgments}
This material is based upon work supported by the National Science Foundation (Platform for the Accelerated Realization, Analysis, and Discovery of Interface Materials (PARADIM)) under Cooperative Agreement No. DMR-2039380. M.D.W., D.A.M., and D.G.S acknoledge support from the National Science Foundation (NSF) under DMR-2122147 
This work made use of the Cornell Center for Materials Research (CCMR) Shared Facilities, which are supported through the NSF MRSEC Program No. DMR1719875. This work made use of a Helios FIB supported by NSF (Grant No. DMR-1539918) and the Cornell Center for Materials Research (CCMR) Shared Facilities, which are supported through the NSF MRSEC Program (Grant No. DMR-1719875). M.D.W. also acknowledges NSD HRD1924204. The authors acknowledge Steve Button for substrate preparation. We gratefully acknowledge Nicholas A. Parker and Yilin Evan Li for help with the AFM experiment, Sankalpa Hazra and Tatiana Kuznetsova for growing the sample used for SIMS experiment, and Dasol Yoon and Xiyue Zhang for providing EDX analysis code.
\end{acknowledgments}
\appendix
\section*{AUTHOR DECLARATIONS}
\textbf{Conflict of Interest}

The authors have no conflicts to disclose.

\section*{Data Availability Statement}
The data that support the findings of this study are available within the article. \textbf{Additional data related to the} film growth and structural characterization by XRD and STEM \textbf{are available at https://doi.org/10.34863/xxxx}

%

%


\bibliography{aipsamp.bib}

\providecommand{\noopsort}[1]{}\providecommand{\singleletter}[1]{#1}%
\begin{thebibliography}{32}%
\makeatletter
\providecommand \@ifxundefined [1]{%
 \@ifx{#1\undefined}
}%
\providecommand \@ifnum [1]{%
 \ifnum #1\expandafter \@firstoftwo
 \else \expandafter \@secondoftwo
 \fi
}%
\providecommand \@ifx [1]{%
 \ifx #1\expandafter \@firstoftwo
 \else \expandafter \@secondoftwo
 \fi
}%
\providecommand \natexlab [1]{#1}%
\providecommand \enquote  [1]{``#1''}%
\providecommand \bibnamefont  [1]{#1}%
\providecommand \bibfnamefont [1]{#1}%
\providecommand \citenamefont [1]{#1}%
\providecommand \href@noop [0]{\@secondoftwo}%
\providecommand \href [0]{\begingroup \@sanitize@url \@href}%
\providecommand \@href[1]{\@@startlink{#1}\@@href}%
\providecommand \@@href[1]{\endgroup#1\@@endlink}%
\providecommand \@sanitize@url [0]{\catcode `\\12\catcode `\$12\catcode
  `\&12\catcode `\#12\catcode `\^12\catcode `\_12\catcode `\%12\relax}%
\providecommand \@@startlink[1]{}%
\providecommand \@@endlink[0]{}%
\providecommand \url  [0]{\begingroup\@sanitize@url \@url }%
\providecommand \@url [1]{\endgroup\@href {#1}{\urlprefix }}%
\providecommand \urlprefix  [0]{URL }%
\providecommand \Eprint [0]{\href }%
\providecommand \doibase [0]{https://doi.org/}%
\providecommand \selectlanguage [0]{\@gobble}%
\providecommand \bibinfo  [0]{\@secondoftwo}%
\providecommand \bibfield  [0]{\@secondoftwo}%
\providecommand \translation [1]{[#1]}%
\providecommand \BibitemOpen [0]{}%
\providecommand \bibitemStop [0]{}%
\providecommand \bibitemNoStop [0]{.\EOS\space}%
\providecommand \EOS [0]{\spacefactor3000\relax}%
\providecommand \BibitemShut  [1]{\csname bibitem#1\endcsname}%
\let\auto@bib@innerbib\@empty
\bibitem [{\citenamefont {Haeni}\ \emph {et~al.}(2004)\citenamefont {Haeni},
  \citenamefont {Irvin}, \citenamefont {Chang}, \citenamefont {Uecker},
  \citenamefont {Reiche}, \citenamefont {Li}, \citenamefont {Choudhury},
  \citenamefont {Tian}, \citenamefont {Hawley}, \citenamefont {Craigo},
  \citenamefont {Tagantsev}, \citenamefont {Pan}, \citenamefont {Streiffer},
  \citenamefont {Chen}, \citenamefont {Kirchoefer}, \citenamefont {Levy},\ and\
  \citenamefont {Schlom}}]{haeni2004room}%
  \BibitemOpen
  \bibfield  {author} {\bibinfo {author} {\bibfnamefont {J.~H.}\ \bibnamefont
  {Haeni}}, \bibinfo {author} {\bibfnamefont {P.}~\bibnamefont {Irvin}},
  \bibinfo {author} {\bibfnamefont {W.}~\bibnamefont {Chang}}, \bibinfo
  {author} {\bibfnamefont {R.}~\bibnamefont {Uecker}}, \bibinfo {author}
  {\bibfnamefont {P.}~\bibnamefont {Reiche}}, \bibinfo {author} {\bibfnamefont
  {Y.~L.}\ \bibnamefont {Li}}, \bibinfo {author} {\bibfnamefont
  {S.}~\bibnamefont {Choudhury}}, \bibinfo {author} {\bibfnamefont
  {W.}~\bibnamefont {Tian}}, \bibinfo {author} {\bibfnamefont {M.~E.}\
  \bibnamefont {Hawley}}, \bibinfo {author} {\bibfnamefont {B.}~\bibnamefont
  {Craigo}}, \bibinfo {author} {\bibfnamefont {A.~K.}\ \bibnamefont
  {Tagantsev}}, \bibinfo {author} {\bibfnamefont {X.~Q.}\ \bibnamefont {Pan}},
  \bibinfo {author} {\bibfnamefont {S.~K.}\ \bibnamefont {Streiffer}}, \bibinfo
  {author} {\bibfnamefont {L.~Q.}\ \bibnamefont {Chen}}, \bibinfo {author}
  {\bibfnamefont {S.~W.}\ \bibnamefont {Kirchoefer}}, \bibinfo {author}
  {\bibfnamefont {J.}~\bibnamefont {Levy}},\ and\ \bibinfo {author}
  {\bibfnamefont {D.~G.}\ \bibnamefont {Schlom}},\ }\bibfield  {title}
  {\enquote {\bibinfo {title} {{Room-temperature ferroelectricity in strained
  \ch{SrTiO3}}},}\ }\href@noop {} {\bibfield  {journal} {\bibinfo  {journal}
  {Nature}\ }\textbf {\bibinfo {volume} {430}},\ \bibinfo {pages} {758--761}
  (\bibinfo {year} {2004})}\BibitemShut {NoStop}%
\bibitem [{\citenamefont {Ahadi}\ \emph {et~al.}(2019)\citenamefont {Ahadi},
  \citenamefont {Galletti}, \citenamefont {Li}, \citenamefont {Salmani-Rezaie},
  \citenamefont {Wu},\ and\ \citenamefont {Stemmer}}]{ahadi2019enhancing}%
  \BibitemOpen
  \bibfield  {author} {\bibinfo {author} {\bibfnamefont {K.}~\bibnamefont
  {Ahadi}}, \bibinfo {author} {\bibfnamefont {L.}~\bibnamefont {Galletti}},
  \bibinfo {author} {\bibfnamefont {Y.}~\bibnamefont {Li}}, \bibinfo {author}
  {\bibfnamefont {S.}~\bibnamefont {Salmani-Rezaie}}, \bibinfo {author}
  {\bibfnamefont {W.}~\bibnamefont {Wu}},\ and\ \bibinfo {author}
  {\bibfnamefont {S.}~\bibnamefont {Stemmer}},\ }\bibfield  {title} {\enquote
  {\bibinfo {title} {{Enhancing superconductivity in \ch{SrTiO3} films with
  strain}},}\ }\href@noop {} {\bibfield  {journal} {\bibinfo  {journal} {Sci.
  Adv.}\ }\textbf {\bibinfo {volume} {5}},\ \bibinfo {pages} {eaaw0120}
  (\bibinfo {year} {2019})}\BibitemShut {NoStop}%
\bibitem [{\citenamefont {Lee}\ \emph {et~al.}(2010)\citenamefont {Lee},
  \citenamefont {Fang}, \citenamefont {Vlahos}, \citenamefont {Ke},
  \citenamefont {Jung}, \citenamefont {Kourkoutis}, \citenamefont {Kim},
  \citenamefont {Ryan}, \citenamefont {Heeg}, \citenamefont {Roeckerath},
  \citenamefont {Goian}, \citenamefont {Bernhagen}, \citenamefont {Uecker},
  \citenamefont {Hammel}, \citenamefont {Rabe}, \citenamefont {Kamba},
  \citenamefont {Schubert}, \citenamefont {Freeland}, \citenamefont {Muller},
  \citenamefont {Fennie}, \citenamefont {Schiffer}, \citenamefont {Gopalan},
  \citenamefont {Johnston-Halperin},\ and\ \citenamefont
  {Schlom}}]{lee2010strong}%
  \BibitemOpen
  \bibfield  {author} {\bibinfo {author} {\bibfnamefont {J.~H.}\ \bibnamefont
  {Lee}}, \bibinfo {author} {\bibfnamefont {L.}~\bibnamefont {Fang}}, \bibinfo
  {author} {\bibfnamefont {E.}~\bibnamefont {Vlahos}}, \bibinfo {author}
  {\bibfnamefont {X.}~\bibnamefont {Ke}}, \bibinfo {author} {\bibfnamefont
  {Y.~W.}\ \bibnamefont {Jung}}, \bibinfo {author} {\bibfnamefont {L.~F.}\
  \bibnamefont {Kourkoutis}}, \bibinfo {author} {\bibfnamefont {J.~W.}\
  \bibnamefont {Kim}}, \bibinfo {author} {\bibfnamefont {P.~J.}\ \bibnamefont
  {Ryan}}, \bibinfo {author} {\bibfnamefont {T.}~\bibnamefont {Heeg}}, \bibinfo
  {author} {\bibfnamefont {M.}~\bibnamefont {Roeckerath}}, \bibinfo {author}
  {\bibfnamefont {V.}~\bibnamefont {Goian}}, \bibinfo {author} {\bibfnamefont
  {M.}~\bibnamefont {Bernhagen}}, \bibinfo {author} {\bibfnamefont
  {R.}~\bibnamefont {Uecker}}, \bibinfo {author} {\bibfnamefont {P.~C.}\
  \bibnamefont {Hammel}}, \bibinfo {author} {\bibfnamefont {K.~M.}\
  \bibnamefont {Rabe}}, \bibinfo {author} {\bibfnamefont {S.}~\bibnamefont
  {Kamba}}, \bibinfo {author} {\bibfnamefont {J.}~\bibnamefont {Schubert}},
  \bibinfo {author} {\bibfnamefont {J.~W.}\ \bibnamefont {Freeland}}, \bibinfo
  {author} {\bibfnamefont {D.~A.}\ \bibnamefont {Muller}}, \bibinfo {author}
  {\bibfnamefont {C.~J.}\ \bibnamefont {Fennie}}, \bibinfo {author}
  {\bibfnamefont {P.}~\bibnamefont {Schiffer}}, \bibinfo {author}
  {\bibfnamefont {V.}~\bibnamefont {Gopalan}}, \bibinfo {author} {\bibfnamefont
  {E.}~\bibnamefont {Johnston-Halperin}},\ and\ \bibinfo {author}
  {\bibfnamefont {D.~G.}\ \bibnamefont {Schlom}},\ }\bibfield  {title}
  {\enquote {\bibinfo {title} {{A strong ferroelectric ferromagnet created by
  means of spin--lattice coupling}},}\ }\href@noop {} {\bibfield  {journal}
  {\bibinfo  {journal} {Nature}\ }\textbf {\bibinfo {volume} {466}},\ \bibinfo
  {pages} {954--958} (\bibinfo {year} {2010})}\BibitemShut {NoStop}%
\bibitem [{\citenamefont {Ahadi}, \citenamefont {Galletti},\ and\ \citenamefont
  {Stemmer}(2017)}]{ahadi2017evidence}%
  \BibitemOpen
  \bibfield  {author} {\bibinfo {author} {\bibfnamefont {K.}~\bibnamefont
  {Ahadi}}, \bibinfo {author} {\bibfnamefont {L.}~\bibnamefont {Galletti}},\
  and\ \bibinfo {author} {\bibfnamefont {S.}~\bibnamefont {Stemmer}},\
  }\bibfield  {title} {\enquote {\bibinfo {title} {{Evidence of a topological
  Hall effect in \ch{Eu_{1-x}Sm_{x}TiO3}}},}\ }\href@noop {} {\bibfield
  {journal} {\bibinfo  {journal} {Appl. Phys. Lett.}\ }\textbf {\bibinfo
  {volume} {111}},\ \bibinfo {pages} {172403} (\bibinfo {year}
  {2017})}\BibitemShut {NoStop}%
\bibitem [{\citenamefont {Ueno}\ \emph {et~al.}(2011)\citenamefont {Ueno},
  \citenamefont {Nakamura}, \citenamefont {Shimotani}, \citenamefont {Yuan},
  \citenamefont {Kimura}, \citenamefont {Nojima}, \citenamefont {Aoki},
  \citenamefont {Iwasa},\ and\ \citenamefont {Kawasaki}}]{ueno2011discovery}%
  \BibitemOpen
  \bibfield  {author} {\bibinfo {author} {\bibfnamefont {K.}~\bibnamefont
  {Ueno}}, \bibinfo {author} {\bibfnamefont {S.}~\bibnamefont {Nakamura}},
  \bibinfo {author} {\bibfnamefont {H.}~\bibnamefont {Shimotani}}, \bibinfo
  {author} {\bibfnamefont {H.}~\bibnamefont {Yuan}}, \bibinfo {author}
  {\bibfnamefont {N.}~\bibnamefont {Kimura}}, \bibinfo {author} {\bibfnamefont
  {T.}~\bibnamefont {Nojima}}, \bibinfo {author} {\bibfnamefont
  {H.}~\bibnamefont {Aoki}}, \bibinfo {author} {\bibfnamefont {Y.}~\bibnamefont
  {Iwasa}},\ and\ \bibinfo {author} {\bibfnamefont {M.}~\bibnamefont
  {Kawasaki}},\ }\bibfield  {title} {\enquote {\bibinfo {title} {{Discovery of
  superconductivity in \ch{KTaO3} by electrostatic carrier doping}},}\
  }\href@noop {} {\bibfield  {journal} {\bibinfo  {journal} {Nat.
  Nanotechnol.}\ }\textbf {\bibinfo {volume} {6}},\ \bibinfo {pages} {408--412}
  (\bibinfo {year} {2011})}\BibitemShut {NoStop}%
\bibitem [{\citenamefont {Liu}\ \emph {et~al.}(2021)\citenamefont {Liu},
  \citenamefont {Yan}, \citenamefont {Jin}, \citenamefont {Ma}, \citenamefont
  {Hsiao}, \citenamefont {Lin}, \citenamefont {Bretz-Sullivan}, \citenamefont
  {Zhou}, \citenamefont {Pearson}, \citenamefont {Fisher}, \citenamefont
  {Jiang}, \citenamefont {Han}, \citenamefont {Zuo}, \citenamefont {Wen},
  \citenamefont {Fong}, \citenamefont {Sun}, \citenamefont {Zhou},\ and\
  \citenamefont {Bhattacharya}}]{changjiang}%
  \BibitemOpen
  \bibfield  {author} {\bibinfo {author} {\bibfnamefont {C.}~\bibnamefont
  {Liu}}, \bibinfo {author} {\bibfnamefont {X.}~\bibnamefont {Yan}}, \bibinfo
  {author} {\bibfnamefont {D.}~\bibnamefont {Jin}}, \bibinfo {author}
  {\bibfnamefont {Y.}~\bibnamefont {Ma}}, \bibinfo {author} {\bibfnamefont
  {H.-W.}\ \bibnamefont {Hsiao}}, \bibinfo {author} {\bibfnamefont
  {Y.}~\bibnamefont {Lin}}, \bibinfo {author} {\bibfnamefont {T.~M.}\
  \bibnamefont {Bretz-Sullivan}}, \bibinfo {author} {\bibfnamefont
  {X.}~\bibnamefont {Zhou}}, \bibinfo {author} {\bibfnamefont {J.}~\bibnamefont
  {Pearson}}, \bibinfo {author} {\bibfnamefont {B.}~\bibnamefont {Fisher}},
  \bibinfo {author} {\bibfnamefont {J.~S.}\ \bibnamefont {Jiang}}, \bibinfo
  {author} {\bibfnamefont {W.}~\bibnamefont {Han}}, \bibinfo {author}
  {\bibfnamefont {J.-M.}\ \bibnamefont {Zuo}}, \bibinfo {author} {\bibfnamefont
  {J.}~\bibnamefont {Wen}}, \bibinfo {author} {\bibfnamefont {D.~D.}\
  \bibnamefont {Fong}}, \bibinfo {author} {\bibfnamefont {J.}~\bibnamefont
  {Sun}}, \bibinfo {author} {\bibfnamefont {H.}~\bibnamefont {Zhou}},\ and\
  \bibinfo {author} {\bibfnamefont {A.}~\bibnamefont {Bhattacharya}},\
  }\bibfield  {title} {\enquote {\bibinfo {title} {{Two-dimensional
  superconductivity and anisotropic transport at \ch{KTaO3} (111)
  interfaces}},}\ }\href@noop {} {\bibfield  {journal} {\bibinfo  {journal}
  {Science}\ }\textbf {\bibinfo {volume} {371}},\ \bibinfo {pages} {716--721}
  (\bibinfo {year} {2021})}\BibitemShut {NoStop}%
\bibitem [{\citenamefont {Gupta}\ \emph {et~al.}(2022)\citenamefont {Gupta},
  \citenamefont {Silotia}, \citenamefont {Kumari}, \citenamefont {Dumen},
  \citenamefont {Goyal}, \citenamefont {Tomar}, \citenamefont {Wadehra},
  \citenamefont {Ayyub},\ and\ \citenamefont {Chakraverty}}]{gupta2022ktao3}%
  \BibitemOpen
  \bibfield  {author} {\bibinfo {author} {\bibfnamefont {A.}~\bibnamefont
  {Gupta}}, \bibinfo {author} {\bibfnamefont {H.}~\bibnamefont {Silotia}},
  \bibinfo {author} {\bibfnamefont {A.}~\bibnamefont {Kumari}}, \bibinfo
  {author} {\bibfnamefont {M.}~\bibnamefont {Dumen}}, \bibinfo {author}
  {\bibfnamefont {S.}~\bibnamefont {Goyal}}, \bibinfo {author} {\bibfnamefont
  {R.}~\bibnamefont {Tomar}}, \bibinfo {author} {\bibfnamefont
  {N.}~\bibnamefont {Wadehra}}, \bibinfo {author} {\bibfnamefont
  {P.}~\bibnamefont {Ayyub}},\ and\ \bibinfo {author} {\bibfnamefont
  {S.}~\bibnamefont {Chakraverty}},\ }\bibfield  {title} {\enquote {\bibinfo
  {title} {{\ch{KTaO3}—The New Kid on the Spintronics Block}},}\ }\href@noop
  {} {\bibfield  {journal} {\bibinfo  {journal} {Adv. Mater.}\ }\textbf
  {\bibinfo {volume} {34}},\ \bibinfo {pages} {2106481} (\bibinfo {year}
  {2022})}\BibitemShut {NoStop}%
\bibitem [{\citenamefont {Al-Tawhid}, \citenamefont {Kumah},\ and\
  \citenamefont {Ahadi}(2021)}]{al2021two}%
  \BibitemOpen
  \bibfield  {author} {\bibinfo {author} {\bibfnamefont {A.~H.}\ \bibnamefont
  {Al-Tawhid}}, \bibinfo {author} {\bibfnamefont {D.~P.}\ \bibnamefont
  {Kumah}},\ and\ \bibinfo {author} {\bibfnamefont {K.}~\bibnamefont {Ahadi}},\
  }\bibfield  {title} {\enquote {\bibinfo {title} {{Two-dimensional electron
  systems and interfacial coupling in \ch{LaCrO3}/\ch{KTaO3}
  heterostructures}},}\ }\href@noop {} {\bibfield  {journal} {\bibinfo
  {journal} {Appl. Phys. Lett.}\ }\textbf {\bibinfo {volume} {118}},\ \bibinfo
  {pages} {192905} (\bibinfo {year} {2021})}\BibitemShut {NoStop}%
\bibitem [{\citenamefont {Al-Tawhid}\ \emph
  {et~al.}(2022{\natexlab{a}})\citenamefont {Al-Tawhid}, \citenamefont
  {Kanter}, \citenamefont {Hatefipour}, \citenamefont {Irving}, \citenamefont
  {Kumah}, \citenamefont {Shabani},\ and\ \citenamefont
  {Ahadi}}]{al2022oxygen}%
  \BibitemOpen
  \bibfield  {author} {\bibinfo {author} {\bibfnamefont {A.~H.}\ \bibnamefont
  {Al-Tawhid}}, \bibinfo {author} {\bibfnamefont {J.}~\bibnamefont {Kanter}},
  \bibinfo {author} {\bibfnamefont {M.}~\bibnamefont {Hatefipour}}, \bibinfo
  {author} {\bibfnamefont {D.~L.}\ \bibnamefont {Irving}}, \bibinfo {author}
  {\bibfnamefont {D.~P.}\ \bibnamefont {Kumah}}, \bibinfo {author}
  {\bibfnamefont {J.}~\bibnamefont {Shabani}},\ and\ \bibinfo {author}
  {\bibfnamefont {K.}~\bibnamefont {Ahadi}},\ }\bibfield  {title} {\enquote
  {\bibinfo {title} {Oxygen vacancy-induced anomalous hall effect in a
  nominally non-magnetic oxide},}\ }\href@noop {} {\bibfield  {journal}
  {\bibinfo  {journal} {Journal of Electronic Materials}\ }\textbf {\bibinfo
  {volume} {51}},\ \bibinfo {pages} {7073--7077} (\bibinfo {year}
  {2022}{\natexlab{a}})}\BibitemShut {NoStop}%
\bibitem [{\citenamefont {Al-Tawhid}\ \emph
  {et~al.}(2022{\natexlab{b}})\citenamefont {Al-Tawhid}, \citenamefont
  {Kanter}, \citenamefont {Hatefipour}, \citenamefont {Kumah}, \citenamefont
  {Shabani},\ and\ \citenamefont {Ahadi}}]{al2022superconductivity}%
  \BibitemOpen
  \bibfield  {author} {\bibinfo {author} {\bibfnamefont {A.~H.}\ \bibnamefont
  {Al-Tawhid}}, \bibinfo {author} {\bibfnamefont {J.}~\bibnamefont {Kanter}},
  \bibinfo {author} {\bibfnamefont {M.}~\bibnamefont {Hatefipour}}, \bibinfo
  {author} {\bibfnamefont {D.~P.}\ \bibnamefont {Kumah}}, \bibinfo {author}
  {\bibfnamefont {J.}~\bibnamefont {Shabani}},\ and\ \bibinfo {author}
  {\bibfnamefont {K.}~\bibnamefont {Ahadi}},\ }\bibfield  {title} {\enquote
  {\bibinfo {title} {Superconductivity and weak anti-localization at ktao3
  (111) interfaces},}\ }\href@noop {} {\bibfield  {journal} {\bibinfo
  {journal} {Journal of Electronic Materials}\ }\textbf {\bibinfo {volume}
  {51}},\ \bibinfo {pages} {6305--6309} (\bibinfo {year}
  {2022}{\natexlab{b}})}\BibitemShut {NoStop}%
\bibitem [{\citenamefont {Nakamura}\ and\ \citenamefont
  {Kimura}(2009)}]{nakamura2009}%
  \BibitemOpen
  \bibfield  {author} {\bibinfo {author} {\bibfnamefont {H.}~\bibnamefont
  {Nakamura}}\ and\ \bibinfo {author} {\bibfnamefont {T.}~\bibnamefont
  {Kimura}},\ }\bibfield  {title} {\enquote {\bibinfo {title} {{Electric field
  tuning of spin-orbit coupling in \ch{KTaO3} field-effect transistors}},}\
  }\href@noop {} {\bibfield  {journal} {\bibinfo  {journal} {Phys. Rev. B}\
  }\textbf {\bibinfo {volume} {80}},\ \bibinfo {pages} {121308} (\bibinfo
  {year} {2009})}\BibitemShut {NoStop}%
\bibitem [{\citenamefont {Wadehra}\ \emph {et~al.}(2020)\citenamefont
  {Wadehra}, \citenamefont {Tomar}, \citenamefont {Varma}, \citenamefont
  {Gopal}, \citenamefont {Singh}, \citenamefont {Dattagupta},\ and\
  \citenamefont {Chakraverty}}]{wadehra2020}%
  \BibitemOpen
  \bibfield  {author} {\bibinfo {author} {\bibfnamefont {N.}~\bibnamefont
  {Wadehra}}, \bibinfo {author} {\bibfnamefont {R.}~\bibnamefont {Tomar}},
  \bibinfo {author} {\bibfnamefont {R.~M.}\ \bibnamefont {Varma}}, \bibinfo
  {author} {\bibfnamefont {R.}~\bibnamefont {Gopal}}, \bibinfo {author}
  {\bibfnamefont {Y.}~\bibnamefont {Singh}}, \bibinfo {author} {\bibfnamefont
  {S.}~\bibnamefont {Dattagupta}},\ and\ \bibinfo {author} {\bibfnamefont
  {S.}~\bibnamefont {Chakraverty}},\ }\bibfield  {title} {\enquote {\bibinfo
  {title} {{Planar Hall effect and anisotropic magnetoresistance in polar-polar
  interface of \ch{LaVO3}-\ch{KTaO3} with strong spin-orbit coupling}},}\
  }\href@noop {} {\bibfield  {journal} {\bibinfo  {journal} {Nat. Commun.}\
  }\textbf {\bibinfo {volume} {11}},\ \bibinfo {pages} {1--7} (\bibinfo {year}
  {2020})}\BibitemShut {NoStop}%
\bibitem [{\citenamefont {Honig}\ and\ \citenamefont
  {Kramer}(1969)}]{honig1969}%
  \BibitemOpen
  \bibfield  {author} {\bibinfo {author} {\bibfnamefont {R.}~\bibnamefont
  {Honig}}\ and\ \bibinfo {author} {\bibfnamefont {D.~A.}\ \bibnamefont
  {Kramer}},\ }\bibfield  {title} {\enquote {\bibinfo {title} {{Vapor pressure
  data for the solid and liquid elements}},}\ }\href@noop {} {\bibfield
  {journal} {\bibinfo  {journal} {RCA Rev.}\ }\textbf {\bibinfo {volume}
  {30}},\ \bibinfo {pages} {285--305} (\bibinfo {year} {1969})}\BibitemShut
  {NoStop}%
\bibitem [{\citenamefont {Gitmans}, \citenamefont {Sitar},\ and\ \citenamefont
  {Günter}(1995)}]{sitar1995}%
  \BibitemOpen
  \bibfield  {author} {\bibinfo {author} {\bibfnamefont {F.}~\bibnamefont
  {Gitmans}}, \bibinfo {author} {\bibfnamefont {Z.}~\bibnamefont {Sitar}},\
  and\ \bibinfo {author} {\bibfnamefont {P.}~\bibnamefont {Günter}},\
  }\bibfield  {title} {\enquote {\bibinfo {title} {{Growth of tantalum oxide
  and lithium tantalate thin films by molecular beam epitaxy}},}\ }\href
  {https://doi.org/https://doi.org/10.1016/0042-207X(95)00077-1} {\bibfield
  {journal} {\bibinfo  {journal} {Vacuum}\ }\textbf {\bibinfo {volume} {46}},\
  \bibinfo {pages} {939--942} (\bibinfo {year} {1995})}\BibitemShut {NoStop}%
\bibitem [{\citenamefont {Barone}\ \emph {et~al.}(2022)\citenamefont {Barone},
  \citenamefont {Foody}, \citenamefont {Hu}, \citenamefont {Sun}, \citenamefont
  {Frye}, \citenamefont {Perera}, \citenamefont {Subedi}, \citenamefont {Paik},
  \citenamefont {Hollin}, \citenamefont {Jeong}, \citenamefont {Lee},
  \citenamefont {Winter}, \citenamefont {Podraza}, \citenamefont {Cho},
  \citenamefont {Hock},\ and\ \citenamefont {Schlom}}]{barone2022}%
  \BibitemOpen
  \bibfield  {author} {\bibinfo {author} {\bibfnamefont {M.}~\bibnamefont
  {Barone}}, \bibinfo {author} {\bibfnamefont {M.}~\bibnamefont {Foody}},
  \bibinfo {author} {\bibfnamefont {Y.}~\bibnamefont {Hu}}, \bibinfo {author}
  {\bibfnamefont {J.}~\bibnamefont {Sun}}, \bibinfo {author} {\bibfnamefont
  {B.}~\bibnamefont {Frye}}, \bibinfo {author} {\bibfnamefont {S.~S.}\
  \bibnamefont {Perera}}, \bibinfo {author} {\bibfnamefont {B.}~\bibnamefont
  {Subedi}}, \bibinfo {author} {\bibfnamefont {H.}~\bibnamefont {Paik}},
  \bibinfo {author} {\bibfnamefont {J.}~\bibnamefont {Hollin}}, \bibinfo
  {author} {\bibfnamefont {M.}~\bibnamefont {Jeong}}, \bibinfo {author}
  {\bibfnamefont {K.}~\bibnamefont {Lee}}, \bibinfo {author} {\bibfnamefont
  {C.~H.}\ \bibnamefont {Winter}}, \bibinfo {author} {\bibfnamefont {N.~J.}\
  \bibnamefont {Podraza}}, \bibinfo {author} {\bibfnamefont {K.}~\bibnamefont
  {Cho}}, \bibinfo {author} {\bibfnamefont {A.}~\bibnamefont {Hock}},\ and\
  \bibinfo {author} {\bibfnamefont {D.~G.}\ \bibnamefont {Schlom}},\ }\bibfield
   {title} {\enquote {\bibinfo {title} {{Growth of \ch{Ta2SnO6} Films, a
  Candidate Wide-Band-Gap p-Type Oxide}},}\ }\href@noop {} {\bibfield
  {journal} {\bibinfo  {journal} {J. Phys. Chem. C}\ }\textbf {\bibinfo
  {volume} {126}},\ \bibinfo {pages} {3764--3775} (\bibinfo {year}
  {2022})}\BibitemShut {NoStop}%
\bibitem [{\citenamefont {Adkison}\ \emph {et~al.}(2020)\citenamefont
  {Adkison}, \citenamefont {Shang}, \citenamefont {Bocklund}, \citenamefont
  {Klimm}, \citenamefont {Schlom},\ and\ \citenamefont {Liu}}]{adkison2020}%
  \BibitemOpen
  \bibfield  {author} {\bibinfo {author} {\bibfnamefont {K.~M.}\ \bibnamefont
  {Adkison}}, \bibinfo {author} {\bibfnamefont {S.-L.}\ \bibnamefont {Shang}},
  \bibinfo {author} {\bibfnamefont {B.~J.}\ \bibnamefont {Bocklund}}, \bibinfo
  {author} {\bibfnamefont {D.}~\bibnamefont {Klimm}}, \bibinfo {author}
  {\bibfnamefont {D.~G.}\ \bibnamefont {Schlom}},\ and\ \bibinfo {author}
  {\bibfnamefont {Z.-K.}\ \bibnamefont {Liu}},\ }\bibfield  {title} {\enquote
  {\bibinfo {title} {{Suitability of binary oxides for molecular-beam epitaxy
  source materials: A comprehensive thermodynamic analysis}},}\ }\href@noop {}
  {\bibfield  {journal} {\bibinfo  {journal} {APL Mater.}\ }\textbf {\bibinfo
  {volume} {8}},\ \bibinfo {pages} {081110} (\bibinfo {year}
  {2020})}\BibitemShut {NoStop}%
\bibitem [{\citenamefont {Du}\ \emph {et~al.}(2020)\citenamefont {Du},
  \citenamefont {Strohbeen}, \citenamefont {Paik}, \citenamefont {Zhang},
  \citenamefont {Genser}, \citenamefont {Rabe}, \citenamefont {Voyles},
  \citenamefont {Schlom},\ and\ \citenamefont {Kawasaki}}]{du2020control}%
  \BibitemOpen
  \bibfield  {author} {\bibinfo {author} {\bibfnamefont {D.}~\bibnamefont
  {Du}}, \bibinfo {author} {\bibfnamefont {P.~J.}\ \bibnamefont {Strohbeen}},
  \bibinfo {author} {\bibfnamefont {H.}~\bibnamefont {Paik}}, \bibinfo {author}
  {\bibfnamefont {C.}~\bibnamefont {Zhang}}, \bibinfo {author} {\bibfnamefont
  {K.~T.}\ \bibnamefont {Genser}}, \bibinfo {author} {\bibfnamefont {K.~M.}\
  \bibnamefont {Rabe}}, \bibinfo {author} {\bibfnamefont {P.~M.}\ \bibnamefont
  {Voyles}}, \bibinfo {author} {\bibfnamefont {D.~G.}\ \bibnamefont {Schlom}},\
  and\ \bibinfo {author} {\bibfnamefont {J.~K.}\ \bibnamefont {Kawasaki}},\
  }\bibfield  {title} {\enquote {\bibinfo {title} {{Control of polymorphism
  during epitaxial growth of hyperferroelectric candidate \ch{LiZnSb} on
  \ch{GaSb} (111) \ch{B}}},}\ }\href@noop {} {\bibfield  {journal} {\bibinfo
  {journal} {J. Vac. Sci. Technol., B: Nanotechnol. Microelectron.: Mater.,
  Process., Meas., Phenom.}\ }\textbf {\bibinfo {volume} {38}},\ \bibinfo
  {pages} {022208} (\bibinfo {year} {2020})}\BibitemShut {NoStop}%
\bibitem [{\citenamefont {Parzyck}\ \emph {et~al.}(2022)\citenamefont
  {Parzyck}, \citenamefont {Galdi}, \citenamefont {Nangoi}, \citenamefont
  {DeBenedetti}, \citenamefont {Balajka}, \citenamefont {Faeth}, \citenamefont
  {Paik}, \citenamefont {Hu}, \citenamefont {Arias}, \citenamefont {Hines},
  \citenamefont {Schlom}, \citenamefont {Shen},\ and\ \citenamefont
  {Maxson}}]{parzyck2022}%
  \BibitemOpen
  \bibfield  {author} {\bibinfo {author} {\bibfnamefont {C.~T.}\ \bibnamefont
  {Parzyck}}, \bibinfo {author} {\bibfnamefont {A.}~\bibnamefont {Galdi}},
  \bibinfo {author} {\bibfnamefont {J.~K.}\ \bibnamefont {Nangoi}}, \bibinfo
  {author} {\bibfnamefont {W.~J.~I.}\ \bibnamefont {DeBenedetti}}, \bibinfo
  {author} {\bibfnamefont {J.}~\bibnamefont {Balajka}}, \bibinfo {author}
  {\bibfnamefont {B.~D.}\ \bibnamefont {Faeth}}, \bibinfo {author}
  {\bibfnamefont {H.}~\bibnamefont {Paik}}, \bibinfo {author} {\bibfnamefont
  {C.}~\bibnamefont {Hu}}, \bibinfo {author} {\bibfnamefont {T.~A.}\
  \bibnamefont {Arias}}, \bibinfo {author} {\bibfnamefont {M.~A.}\ \bibnamefont
  {Hines}}, \bibinfo {author} {\bibfnamefont {D.~G.}\ \bibnamefont {Schlom}},
  \bibinfo {author} {\bibfnamefont {K.~M.}\ \bibnamefont {Shen}},\ and\
  \bibinfo {author} {\bibfnamefont {J.~M.}\ \bibnamefont {Maxson}},\ }\bibfield
   {title} {\enquote {\bibinfo {title} {{Single-Crystal Alkali Antimonide
  Photocathodes: High Efficiency in the Ultrathin Limit}},}\ }\href@noop {}
  {\bibfield  {journal} {\bibinfo  {journal} {Phys. Rev. Lett.}\ }\textbf
  {\bibinfo {volume} {128}},\ \bibinfo {pages} {114801} (\bibinfo {year}
  {2022})}\BibitemShut {NoStop}%
\bibitem [{\citenamefont {Okamoto}(1992)}]{okamoto1992k}%
  \BibitemOpen
  \bibfield  {author} {\bibinfo {author} {\bibfnamefont {H.}~\bibnamefont
  {Okamoto}},\ }\bibfield  {title} {\enquote {\bibinfo {title} {{In-K
  (Indium-Potassium)}},}\ }\href@noop {} {\bibfield  {journal} {\bibinfo
  {journal} {J. Phase Equilib.}\ }\textbf {\bibinfo {volume} {13}},\ \bibinfo
  {pages} {217--218} (\bibinfo {year} {1992})}\BibitemShut {NoStop}%
\bibitem [{\citenamefont {Koster}\ \emph {et~al.}(1998)\citenamefont {Koster},
  \citenamefont {Kropman}, \citenamefont {Rijnders}, \citenamefont {Blank},\
  and\ \citenamefont {Rogalla}}]{koster1998}%
  \BibitemOpen
  \bibfield  {author} {\bibinfo {author} {\bibfnamefont {G.}~\bibnamefont
  {Koster}}, \bibinfo {author} {\bibfnamefont {B.~L.}\ \bibnamefont {Kropman}},
  \bibinfo {author} {\bibfnamefont {G.~J.}\ \bibnamefont {Rijnders}}, \bibinfo
  {author} {\bibfnamefont {D.~H.}\ \bibnamefont {Blank}},\ and\ \bibinfo
  {author} {\bibfnamefont {H.}~\bibnamefont {Rogalla}},\ }\bibfield  {title}
  {\enquote {\bibinfo {title} {{Quasi-ideal strontium titanate crystal surfaces
  through formation of strontium hydroxide}},}\ }\href@noop {} {\bibfield
  {journal} {\bibinfo  {journal} {Appl. Phys. Lett.}\ }\textbf {\bibinfo
  {volume} {73}},\ \bibinfo {pages} {2920--2922} (\bibinfo {year}
  {1998})}\BibitemShut {NoStop}%
\bibitem [{\citenamefont {Zhurova}\ \emph {et~al.}(2000)\citenamefont
  {Zhurova}, \citenamefont {Ivanov}, \citenamefont {Zavodnik},\ and\
  \citenamefont {Tsirelson}}]{zhurova2000}%
  \BibitemOpen
  \bibfield  {author} {\bibinfo {author} {\bibfnamefont {E.~A.}\ \bibnamefont
  {Zhurova}}, \bibinfo {author} {\bibfnamefont {Y.}~\bibnamefont {Ivanov}},
  \bibinfo {author} {\bibfnamefont {V.}~\bibnamefont {Zavodnik}},\ and\
  \bibinfo {author} {\bibfnamefont {V.}~\bibnamefont {Tsirelson}},\ }\bibfield
  {title} {\enquote {\bibinfo {title} {{Electron density and atomic
  displacements in \ch{KTaO3}}},}\ }\href@noop {} {\bibfield  {journal}
  {\bibinfo  {journal} {Acta Crystallogr., Sect. B: Struct. Sci.}\ }\textbf
  {\bibinfo {volume} {56}},\ \bibinfo {pages} {594--600} (\bibinfo {year}
  {2000})}\BibitemShut {NoStop}%
\bibitem [{\citenamefont {Uecker}\ \emph {et~al.}(2008)\citenamefont {Uecker},
  \citenamefont {Velickov}, \citenamefont {Klimm}, \citenamefont {Bertram},
  \citenamefont {Bernhagen}, \citenamefont {Rabe}, \citenamefont {Albrecht},
  \citenamefont {Fornari},\ and\ \citenamefont {Schlom}}]{uecker2008}%
  \BibitemOpen
  \bibfield  {author} {\bibinfo {author} {\bibfnamefont {R.}~\bibnamefont
  {Uecker}}, \bibinfo {author} {\bibfnamefont {B.}~\bibnamefont {Velickov}},
  \bibinfo {author} {\bibfnamefont {D.}~\bibnamefont {Klimm}}, \bibinfo
  {author} {\bibfnamefont {R.}~\bibnamefont {Bertram}}, \bibinfo {author}
  {\bibfnamefont {M.}~\bibnamefont {Bernhagen}}, \bibinfo {author}
  {\bibfnamefont {M.}~\bibnamefont {Rabe}}, \bibinfo {author} {\bibfnamefont
  {M.}~\bibnamefont {Albrecht}}, \bibinfo {author} {\bibfnamefont
  {R.}~\bibnamefont {Fornari}},\ and\ \bibinfo {author} {\bibfnamefont
  {D.}~\bibnamefont {Schlom}},\ }\bibfield  {title} {\enquote {\bibinfo {title}
  {{Properties of rare-earth scandate single crystals (\ch{Re}=
  \ch{Nd}-\ch{Dy})}},}\ }\href@noop {} {\bibfield  {journal} {\bibinfo
  {journal} {J. Cryst. Growth}\ }\textbf {\bibinfo {volume} {310}},\ \bibinfo
  {pages} {2649--2658} (\bibinfo {year} {2008})}\BibitemShut {NoStop}%
\bibitem [{\citenamefont {Culbertson}\ \emph {et~al.}(2020)\citenamefont
  {Culbertson}, \citenamefont {Flak}, \citenamefont {Yatskin}, \citenamefont
  {Cheong}, \citenamefont {Cann},\ and\ \citenamefont
  {Dolgos}}]{culbertson2020}%
  \BibitemOpen
  \bibfield  {author} {\bibinfo {author} {\bibfnamefont {C.~M.}\ \bibnamefont
  {Culbertson}}, \bibinfo {author} {\bibfnamefont {A.~T.}\ \bibnamefont
  {Flak}}, \bibinfo {author} {\bibfnamefont {M.}~\bibnamefont {Yatskin}},
  \bibinfo {author} {\bibfnamefont {P.~H.-Y.}\ \bibnamefont {Cheong}}, \bibinfo
  {author} {\bibfnamefont {D.~P.}\ \bibnamefont {Cann}},\ and\ \bibinfo
  {author} {\bibfnamefont {M.~R.}\ \bibnamefont {Dolgos}},\ }\bibfield  {title}
  {\enquote {\bibinfo {title} {{Neutron total scattering studies of group II
  titanates (\ch{ATiO3}, \ch{A^{2+}}= \ch{Mg}, \ch{Ca}, \ch{Sr},\ch{Ba}) }},}\
  }\href@noop {} {\bibfield  {journal} {\bibinfo  {journal} {Sci. Rep.}\
  }\textbf {\bibinfo {volume} {10}},\ \bibinfo {pages} {3729} (\bibinfo {year}
  {2020})}\BibitemShut {NoStop}%
\bibitem [{\citenamefont {Madelung}(1990)}]{LandoltBornstein1990}%
  \BibitemOpen
  \bibinfo {editor} {\bibfnamefont {O.}~\bibnamefont {Madelung}},\ ed.,\
  \href@noop {} {\emph {\bibinfo {title} {{Landolt-Börnstein: Numerical Data
  and Functional Relationsships in Science and Technology}}}},\ Vol.\ \bibinfo
  {volume} {29a: Low frequency Properties of Dielectric Crystals}\ (\bibinfo
  {publisher} {Springer-Verlag Berlin Heidelberg},\ \bibinfo {year} {1990})\
  p.~\bibinfo {pages} {96}\BibitemShut {NoStop}%
\bibitem [{\citenamefont {Uecker}\ \emph {et~al.}(2013)\citenamefont {Uecker},
  \citenamefont {Klimm}, \citenamefont {Bertram}, \citenamefont {Bernhagen},
  \citenamefont {Schulze-Jonack}, \citenamefont {Brützam}, \citenamefont
  {Kwasniewski}, \citenamefont {Gesing},\ and\ \citenamefont
  {Schlom}}]{uecker2013}%
  \BibitemOpen
  \bibfield  {author} {\bibinfo {author} {\bibfnamefont {R.}~\bibnamefont
  {Uecker}}, \bibinfo {author} {\bibfnamefont {D.}~\bibnamefont {Klimm}},
  \bibinfo {author} {\bibfnamefont {R.}~\bibnamefont {Bertram}}, \bibinfo
  {author} {\bibfnamefont {M.}~\bibnamefont {Bernhagen}}, \bibinfo {author}
  {\bibfnamefont {I.}~\bibnamefont {Schulze-Jonack}}, \bibinfo {author}
  {\bibfnamefont {M.}~\bibnamefont {Brützam}}, \bibinfo {author}
  {\bibfnamefont {A.}~\bibnamefont {Kwasniewski}}, \bibinfo {author}
  {\bibfnamefont {T.}~\bibnamefont {Gesing}},\ and\ \bibinfo {author}
  {\bibfnamefont {D.}~\bibnamefont {Schlom}},\ }\bibfield  {title} {\enquote
  {\bibinfo {title} {{Growth and Investigation of \ch{Nd_{1-x}Sm_{x}ScO3} and
  \ch{Sm_{1-x}Gd_{x}ScO3} Solid-Solution Single Crystals}},}\ }\href@noop {}
  {\bibfield  {journal} {\bibinfo  {journal} {Acta Phys. Pol., A}\ }\textbf
  {\bibinfo {volume} {2}},\ \bibinfo {pages} {295--300} (\bibinfo {year}
  {2013})}\BibitemShut {NoStop}%
\bibitem [{\citenamefont {Brooks}\ \emph {et~al.}(2009)\citenamefont {Brooks},
  \citenamefont {Kourkoutis}, \citenamefont {Heeg}, \citenamefont {Schubert},
  \citenamefont {Muller},\ and\ \citenamefont {Schlom}}]{brooks2009}%
  \BibitemOpen
  \bibfield  {author} {\bibinfo {author} {\bibfnamefont {C.}~\bibnamefont
  {Brooks}}, \bibinfo {author} {\bibfnamefont {L.~F.}\ \bibnamefont
  {Kourkoutis}}, \bibinfo {author} {\bibfnamefont {T.}~\bibnamefont {Heeg}},
  \bibinfo {author} {\bibfnamefont {J.}~\bibnamefont {Schubert}}, \bibinfo
  {author} {\bibfnamefont {D.}~\bibnamefont {Muller}},\ and\ \bibinfo {author}
  {\bibfnamefont {D.}~\bibnamefont {Schlom}},\ }\bibfield  {title} {\enquote
  {\bibinfo {title} {{Growth of homoepitaxial \ch{SrTiO3} thin films by
  molecular-beam epitaxy}},}\ }\href@noop {} {\bibfield  {journal} {\bibinfo
  {journal} {Appl. Phys. Lett.}\ }\textbf {\bibinfo {volume} {94}},\ \bibinfo
  {pages} {162905} (\bibinfo {year} {2009})}\BibitemShut {NoStop}%
\bibitem [{\citenamefont {Jalan}, \citenamefont {Moetakef},\ and\ \citenamefont
  {Stemmer}(2009)}]{jalan2009}%
  \BibitemOpen
  \bibfield  {author} {\bibinfo {author} {\bibfnamefont {B.}~\bibnamefont
  {Jalan}}, \bibinfo {author} {\bibfnamefont {P.}~\bibnamefont {Moetakef}},\
  and\ \bibinfo {author} {\bibfnamefont {S.}~\bibnamefont {Stemmer}},\
  }\bibfield  {title} {\enquote {\bibinfo {title} {{Molecular beam epitaxy of
  \ch{SrTiO3} with a growth window}},}\ }\href@noop {} {\bibfield  {journal}
  {\bibinfo  {journal} {Appl. Phys. Lett.}\ }\textbf {\bibinfo {volume} {95}},\
  \bibinfo {pages} {032906} (\bibinfo {year} {2009})}\BibitemShut {NoStop}%
\bibitem [{\citenamefont {Haislmaier}\ \emph {et~al.}(2016)\citenamefont
  {Haislmaier}, \citenamefont {Grimley}, \citenamefont {Biegalski},
  \citenamefont {LeBeau}, \citenamefont {Trolier-McKinstry}, \citenamefont
  {Gopalan},\ and\ \citenamefont {Engel-Herbert}}]{haislmaier2016}%
  \BibitemOpen
  \bibfield  {author} {\bibinfo {author} {\bibfnamefont {R.~C.}\ \bibnamefont
  {Haislmaier}}, \bibinfo {author} {\bibfnamefont {E.~D.}\ \bibnamefont
  {Grimley}}, \bibinfo {author} {\bibfnamefont {M.~D.}\ \bibnamefont
  {Biegalski}}, \bibinfo {author} {\bibfnamefont {J.~M.}\ \bibnamefont
  {LeBeau}}, \bibinfo {author} {\bibfnamefont {S.}~\bibnamefont
  {Trolier-McKinstry}}, \bibinfo {author} {\bibfnamefont {V.}~\bibnamefont
  {Gopalan}},\ and\ \bibinfo {author} {\bibfnamefont {R.}~\bibnamefont
  {Engel-Herbert}},\ }\bibfield  {title} {\enquote {\bibinfo {title}
  {{Unleashing strain induced ferroelectricity in complex oxide thin films via
  precise stoichiometry control}},}\ }\href@noop {} {\bibfield  {journal}
  {\bibinfo  {journal} {Adv. Funct. Mater.}\ }\textbf {\bibinfo {volume}
  {26}},\ \bibinfo {pages} {7271--7279} (\bibinfo {year} {2016})}\BibitemShut
  {NoStop}%
\bibitem [{\citenamefont {Moyer}, \citenamefont {Eaton},\ and\ \citenamefont
  {Engel-Herbert}(2013)}]{moyer2013}%
  \BibitemOpen
  \bibfield  {author} {\bibinfo {author} {\bibfnamefont {J.~A.}\ \bibnamefont
  {Moyer}}, \bibinfo {author} {\bibfnamefont {C.}~\bibnamefont {Eaton}},\ and\
  \bibinfo {author} {\bibfnamefont {R.}~\bibnamefont {Engel-Herbert}},\
  }\bibfield  {title} {\enquote {\bibinfo {title} {{Highly conductive
  \ch{SrVO3} as a bottom electrode for functional perovskite oxides}},}\
  }\href@noop {} {\bibfield  {journal} {\bibinfo  {journal} {Adv. Mater.}\
  }\textbf {\bibinfo {volume} {25}},\ \bibinfo {pages} {3578--3582} (\bibinfo
  {year} {2013})}\BibitemShut {NoStop}%
\bibitem [{\citenamefont {Zhang}\ \emph {et~al.}(2015)\citenamefont {Zhang},
  \citenamefont {Dedon}, \citenamefont {Martin},\ and\ \citenamefont
  {Engel-Herbert}}]{zhang2015}%
  \BibitemOpen
  \bibfield  {author} {\bibinfo {author} {\bibfnamefont {H.-T.}\ \bibnamefont
  {Zhang}}, \bibinfo {author} {\bibfnamefont {L.~R.}\ \bibnamefont {Dedon}},
  \bibinfo {author} {\bibfnamefont {L.~W.}\ \bibnamefont {Martin}},\ and\
  \bibinfo {author} {\bibfnamefont {R.}~\bibnamefont {Engel-Herbert}},\
  }\bibfield  {title} {\enquote {\bibinfo {title} {{Self-regulated growth of
  \ch{LaVO3} thin films by hybrid molecular beam epitaxy}},}\ }\href@noop {}
  {\bibfield  {journal} {\bibinfo  {journal} {Appl. Phys. Lett.}\ }\textbf
  {\bibinfo {volume} {106}},\ \bibinfo {pages} {233102} (\bibinfo {year}
  {2015})}\BibitemShut {NoStop}%
\bibitem [{\citenamefont {Tyunina}\ \emph {et~al.}(2010)\citenamefont
  {Tyunina}, \citenamefont {Narkilahti}, \citenamefont {Plekh}, \citenamefont
  {Oja}, \citenamefont {Nieminen}, \citenamefont {Dejneka},\ and\ \citenamefont
  {Trepakov}}]{tyunina2010}%
  \BibitemOpen
  \bibfield  {author} {\bibinfo {author} {\bibfnamefont {M.}~\bibnamefont
  {Tyunina}}, \bibinfo {author} {\bibfnamefont {J.}~\bibnamefont {Narkilahti}},
  \bibinfo {author} {\bibfnamefont {M.}~\bibnamefont {Plekh}}, \bibinfo
  {author} {\bibfnamefont {R.}~\bibnamefont {Oja}}, \bibinfo {author}
  {\bibfnamefont {R.~M.}\ \bibnamefont {Nieminen}}, \bibinfo {author}
  {\bibfnamefont {A.}~\bibnamefont {Dejneka}},\ and\ \bibinfo {author}
  {\bibfnamefont {V.}~\bibnamefont {Trepakov}},\ }\bibfield  {title} {\enquote
  {\bibinfo {title} {{Evidence for strain-induced ferroelectric order in
  epitaxial thin-film \ch{KTaO3}}},}\ }\href@noop {} {\bibfield  {journal}
  {\bibinfo  {journal} {Phys. Rev. Lett.}\ }\textbf {\bibinfo {volume} {104}},\
  \bibinfo {pages} {227601} (\bibinfo {year} {2010})}\BibitemShut {NoStop}%
\bibitem [{\citenamefont {Thompson}\ \emph {et~al.}(2014)\citenamefont
  {Thompson}, \citenamefont {Hwang}, \citenamefont {Nichols}, \citenamefont
  {Connell}, \citenamefont {Stemmer},\ and\ \citenamefont
  {Seo}}]{thompson2014}%
  \BibitemOpen
  \bibfield  {author} {\bibinfo {author} {\bibfnamefont {J.}~\bibnamefont
  {Thompson}}, \bibinfo {author} {\bibfnamefont {J.}~\bibnamefont {Hwang}},
  \bibinfo {author} {\bibfnamefont {J.}~\bibnamefont {Nichols}}, \bibinfo
  {author} {\bibfnamefont {J.~G.}\ \bibnamefont {Connell}}, \bibinfo {author}
  {\bibfnamefont {S.}~\bibnamefont {Stemmer}},\ and\ \bibinfo {author}
  {\bibfnamefont {S.~S.~A.}\ \bibnamefont {Seo}},\ }\bibfield  {title}
  {\enquote {\bibinfo {title} {{Alleviating polarity-conflict at the
  heterointerfaces of \ch{KTaO3}/\ch{GdScO3} polar complex-oxides}},}\
  }\href@noop {} {\bibfield  {journal} {\bibinfo  {journal} {Appl. Phys.
  Lett.}\ }\textbf {\bibinfo {volume} {105}},\ \bibinfo {pages} {102901}
  (\bibinfo {year} {2014})}\BibitemShut {NoStop}%
\end{thebibliography}%


\providecommand{\noopsort}[1]{}\providecommand{\singleletter}[1]{#1}%
\begin{thebibliography}{7}%
\makeatletter
\providecommand \@ifxundefined [1]{%
 \@ifx{#1\undefined}
}%
\providecommand \@ifnum [1]{%
 \ifnum #1\expandafter \@firstoftwo
 \else \expandafter \@secondoftwo
 \fi
}%
\providecommand \@ifx [1]{%
 \ifx #1\expandafter \@firstoftwo
 \else \expandafter \@secondoftwo
 \fi
}%
\providecommand \natexlab [1]{#1}%
\providecommand \enquote  [1]{``#1''}%
\providecommand \bibnamefont  [1]{#1}%
\providecommand \bibfnamefont [1]{#1}%
\providecommand \citenamefont [1]{#1}%
\providecommand \href@noop [0]{\@secondoftwo}%
\providecommand \href [0]{\begingroup \@sanitize@url \@href}%
\providecommand \@href[1]{\@@startlink{#1}\@@href}%
\providecommand \@@href[1]{\endgroup#1\@@endlink}%
\providecommand \@sanitize@url [0]{\catcode `\\12\catcode `\$12\catcode
  `\&12\catcode `\#12\catcode `\^12\catcode `\_12\catcode `\%12\relax}%
\providecommand \@@startlink[1]{}%
\providecommand \@@endlink[0]{}%
\providecommand \url  [0]{\begingroup\@sanitize@url \@url }%
\providecommand \@url [1]{\endgroup\@href {#1}{\urlprefix }}%
\providecommand \urlprefix  [0]{URL }%
\providecommand \Eprint [0]{\href }%
\providecommand \doibase [0]{https://doi.org/}%
\providecommand \selectlanguage [0]{\@gobble}%
\providecommand \bibinfo  [0]{\@secondoftwo}%
\providecommand \bibfield  [0]{\@secondoftwo}%
\providecommand \translation [1]{[#1]}%
\providecommand \BibitemOpen [0]{}%
\providecommand \bibitemStop [0]{}%
\providecommand \bibitemNoStop [0]{.\EOS\space}%
\providecommand \EOS [0]{\spacefactor3000\relax}%
\providecommand \BibitemShut  [1]{\csname bibitem#1\endcsname}%
\let\auto@bib@innerbib\@empty
\bibitem [{\citenamefont {Honig}\ and\ \citenamefont
  {Kramer}(1969)}]{honig1969}%
  \BibitemOpen
  \bibfield  {author} {\bibinfo {author} {\bibfnamefont {R.}~\bibnamefont
  {Honig}}\ and\ \bibinfo {author} {\bibfnamefont {D.~A.}\ \bibnamefont
  {Kramer}},\ }\bibfield  {title} {\enquote {\bibinfo {title} {{Vapor pressure
  data for the solid and liquid elements}},}\ }\href@noop {} {\bibfield
  {journal} {\bibinfo  {journal} {RCA Rev.}\ }\textbf {\bibinfo {volume}
  {30}},\ \bibinfo {pages} {285--305} (\bibinfo {year} {1969})}\BibitemShut
  {NoStop}%
\bibitem [{\citenamefont {Loeb}(1934)}]{Loeb}%
  \BibitemOpen
  \bibfield  {author} {\bibinfo {author} {\bibfnamefont {L.}~\bibnamefont
  {Loeb}},\ }\href@noop {} {\emph {\bibinfo {title} {The Kinetic Theory of
  Gases: Being a Text and Reference Book Whose Purpose is to Combine the
  Classical Deductions with Recent Experimental Advances in a Convenient Form
  for Student and Investigator}}},\ \bibinfo {edition} {2nd}\ ed.\ (\bibinfo
  {publisher} {McGraw-Hill, New York},\ \bibinfo {year} {1934})\ pp.\ \bibinfo
  {pages} {19, 106}\BibitemShut {NoStop}%
\bibitem [{\citenamefont {Nair}\ \emph {et~al.}(2018)\citenamefont {Nair},
  \citenamefont {Liu}, \citenamefont {Ruf}, \citenamefont {Schreiber},
  \citenamefont {Shang}, \citenamefont {Baek}, \citenamefont {Goodge},
  \citenamefont {Kourkoutis}, \citenamefont {Liu}, \citenamefont {Shen},\ and\
  \citenamefont {Schlom}}]{Nair2018}%
  \BibitemOpen
  \bibfield  {author} {\bibinfo {author} {\bibfnamefont {H.~P.}\ \bibnamefont
  {Nair}}, \bibinfo {author} {\bibfnamefont {Y.}~\bibnamefont {Liu}}, \bibinfo
  {author} {\bibfnamefont {J.~P.}\ \bibnamefont {Ruf}}, \bibinfo {author}
  {\bibfnamefont {N.~J.}\ \bibnamefont {Schreiber}}, \bibinfo {author}
  {\bibfnamefont {S.-L.}\ \bibnamefont {Shang}}, \bibinfo {author}
  {\bibfnamefont {D.~J.}\ \bibnamefont {Baek}}, \bibinfo {author}
  {\bibfnamefont {B.~H.}\ \bibnamefont {Goodge}}, \bibinfo {author}
  {\bibfnamefont {L.~F.}\ \bibnamefont {Kourkoutis}}, \bibinfo {author}
  {\bibfnamefont {Z.-K.}\ \bibnamefont {Liu}}, \bibinfo {author} {\bibfnamefont
  {K.~M.}\ \bibnamefont {Shen}},\ and\ \bibinfo {author} {\bibfnamefont
  {D.~G.}\ \bibnamefont {Schlom}},\ }\bibfield  {title} {\enquote {\bibinfo
  {title} {{Synthesis science of \ch{SrRuO3} and \ch{CaRuO3} epitaxial films
  with high residual resistivity ratios}},}\ }\href@noop {} {\bibfield
  {journal} {\bibinfo  {journal} {APL Mater.}\ }\textbf {\bibinfo {volume}
  {6}},\ \bibinfo {pages} {046101} (\bibinfo {year} {2018})}\BibitemShut
  {NoStop}%
\bibitem [{\citenamefont {Lamoreaux}\ and\ \citenamefont
  {Hildenbrand}(1984)}]{lamoreaux1984}%
  \BibitemOpen
  \bibfield  {author} {\bibinfo {author} {\bibfnamefont {R.}~\bibnamefont
  {Lamoreaux}}\ and\ \bibinfo {author} {\bibfnamefont {D.}~\bibnamefont
  {Hildenbrand}},\ }\bibfield  {title} {\enquote {\bibinfo {title} {{High
  temperature vaporization behavior of oxides. I. Alkali metal binary
  oxides}},}\ }\href@noop {} {\bibfield  {journal} {\bibinfo  {journal} {J.
  Phys. Chem. Ref. Data}\ }\textbf {\bibinfo {volume} {13}},\ \bibinfo {pages}
  {151--173} (\bibinfo {year} {1984})}\BibitemShut {NoStop}%
\bibitem [{\citenamefont {Tomar}\ \emph {et~al.}(2018)\citenamefont {Tomar},
  \citenamefont {Wadehra}, \citenamefont {Budhiraja}, \citenamefont {Prakash},\
  and\ \citenamefont {Chakraverty}}]{tomar2018}%
  \BibitemOpen
  \bibfield  {author} {\bibinfo {author} {\bibfnamefont {R.}~\bibnamefont
  {Tomar}}, \bibinfo {author} {\bibfnamefont {N.}~\bibnamefont {Wadehra}},
  \bibinfo {author} {\bibfnamefont {V.}~\bibnamefont {Budhiraja}}, \bibinfo
  {author} {\bibfnamefont {B.}~\bibnamefont {Prakash}},\ and\ \bibinfo {author}
  {\bibfnamefont {S.}~\bibnamefont {Chakraverty}},\ }\bibfield  {title}
  {\enquote {\bibinfo {title} {{Realization of single terminated surface of
  perovskite oxide single crystals and their band
  profile:(\ch{LaAlO3})$_{0.3}$(\ch{Sr2AlTaO6})$_{0.7}$, \ch{SrTiO3} and
  \ch{KTaO3} case study}},}\ }\href@noop {} {\bibfield  {journal} {\bibinfo
  {journal} {Appl. Surf. Sci.}\ }\textbf {\bibinfo {volume} {427}},\ \bibinfo
  {pages} {861--866} (\bibinfo {year} {2018})}\BibitemShut {NoStop}%
\bibitem [{\citenamefont {Stevie}(2015)}]{stevie2015secondary}%
  \BibitemOpen
  \bibfield  {author} {\bibinfo {author} {\bibfnamefont {F.}~\bibnamefont
  {Stevie}},\ }\href@noop {} {\emph {\bibinfo {title} {Secondary ion mass
  spectrometry: applications for depth profiling and surface
  characterization}}}\ (\bibinfo  {publisher} {Momentum Press},\ \bibinfo
  {year} {2015})\BibitemShut {NoStop}%
\bibitem [{\citenamefont
  {{https://www.eag.com/wp-content/uploads/2022/09/M-065922\_SIMS-Detection-Limits\_Ga2O3.pdf}}()}]{eag}%
  \BibitemOpen
  \bibfield  {author} {\bibinfo {author} {\bibnamefont
  {{https://www.eag.com/wp-content/uploads/2022/09/M-065922\_SIMS-Detection-Limits\_Ga2O3.pdf}}},\
  }\href@noop {} {}\BibitemShut {NoStop}%
\end{thebibliography}%

\end{document}



\title[Supplementary Material 
]{Supplementary Material Molecular Beam Epitaxy of \ch{KTaO3}} 

\author{Tobias Schwaigert}
\affiliation{Platform for the Accelerated Realization, Analysis, and Discovery of Interface Materials (PARADIM), Cornell University, Ithaca, New York 14853, USA}
\affiliation{Department of Materials Science and Engineering, Cornell University, Ithaca, NY 14853, USA}
\author{Salva Salmani-Rezaie}
\affiliation{School of Applied and Engineering Physics, Cornell University, Ithaca, New York 14853, USA}
\affiliation{Kavli Institute at Cornell for Nanoscale Science, Cornell University, Ithaca, New York 14853, USA}
\author{Matthew R. Barone}
\affiliation{Platform for the Accelerated Realization, Analysis, and Discovery of Interface Materials (PARADIM), Cornell University, Ithaca, New York 14853, USA}
\affiliation{Department of Materials Science and Engineering, Cornell University, Ithaca, NY 14853, USA}
\author{Hanjong Paik}
\affiliation{Platform for the Accelerated Realization, Analysis, and Discovery of Interface Materials (PARADIM), Cornell University, Ithaca, New York 14853, USA}
\affiliation{School of Electrical \& Computer Engineering, University of Oklahoma, Norman, OK 73019, USA}
\affiliation{Center for Quantum Research and Technology, University of Oklahoma, Norman, OK 73019, USA}
\author{Ethan Ray}
\affiliation{Platform for the Accelerated Realization, Analysis, and Discovery of Interface Materials (PARADIM), Cornell University, Ithaca, New York 14853, USA}
\author{Michael D. Williams}
\affiliation{Department of Physics, Clark Atlanta University, Atlanta, Georgia 30314, USA}
\author{David A. Muller}
\affiliation{School of Applied and Engineering Physics, Cornell University, Ithaca, New York 14853, USA}
\affiliation{Kavli Institute at Cornell for Nanoscale Science, Cornell University, Ithaca, New York 14853, USA}
\author{Darrell G. Schlom}
\affiliation{Platform for the Accelerated Realization, Analysis, and Discovery of Interface Materials (PARADIM), Cornell University, Ithaca, New York 14853, USA}
\affiliation{Department of Materials Science and Engineering, Cornell University, Ithaca, NY 14853, USA}
\affiliation{Kavli Institute at Cornell for Nanoscale Science, Cornell University, Ithaca, New York 14853, USA}
\affiliation{Leibniz-Institut für Kristallzüchtung, Max-Born-Str. 2, 12489 Berlin, Germany}
\author{Kaveh Ahadi}
\email[kahadi@ncsu.edu]{}
\affiliation{Department of Materials Science and Engineering, North Carolina State University, Raleigh, NC 27265 USA}
\affiliation{Department of Physics, North Carolina State University, Raleigh, North Carolina 27695, USA}
\date{\today}
\maketitle 
\section{Using the QCM to measure the potassium flux}
The considerable vapor pressure of potassium should be considered when making QCM measurements.The vapor pressure, $P_i$, of potassium  at room temperature (the temperature of the QCM during the potassium flux measurement) is 1.8$\times 10^{-8}$ Torr.\cite{honig1969} From the kinetic theory of gases, the maximum desorption rate (corresponding to an evaporation coefficient $\alpha$=1) from the QCM at room temperature can be calculated from this vapor pressure using the equation:\cite{Loeb}
\begin{equation}
    \Phi_K = \frac{P_i}{\sqrt{2 \pi m_K k_B T}}
\end{equation}
\begin{itemize}
  \item $k_B$ is the Boltzmann constant
  \item $m_K$ is the atomic mass of potassium
  \item $T$ is the tempature of the potassium atoms (the temperature of the QCM, i.e., 300 K)
\end{itemize}
The resulting flux emanating from the QCM is $\Phi_{leaving} \sim$ 5.8$\times 10^{12}$ atoms/$\text{cm}^2$/s. Considering that the potassium flux measured by the QCM is $\Phi \sim$ (4-7)$\times 10^{13}$ atoms/$\text{cm}^2$/s, the flux of potassium arriving at the QCM is an order of magnitude higher than what is leaving. We are thus measuring roughly 10 \% less flux then we are actually supplying due to loss from evaporation. This is within the error of the QCM.
\section{Should accumulation of potassium oxide on the film surface be expected?}
\begin{figure}[H]
\centering
\includegraphics[width=\textwidth]{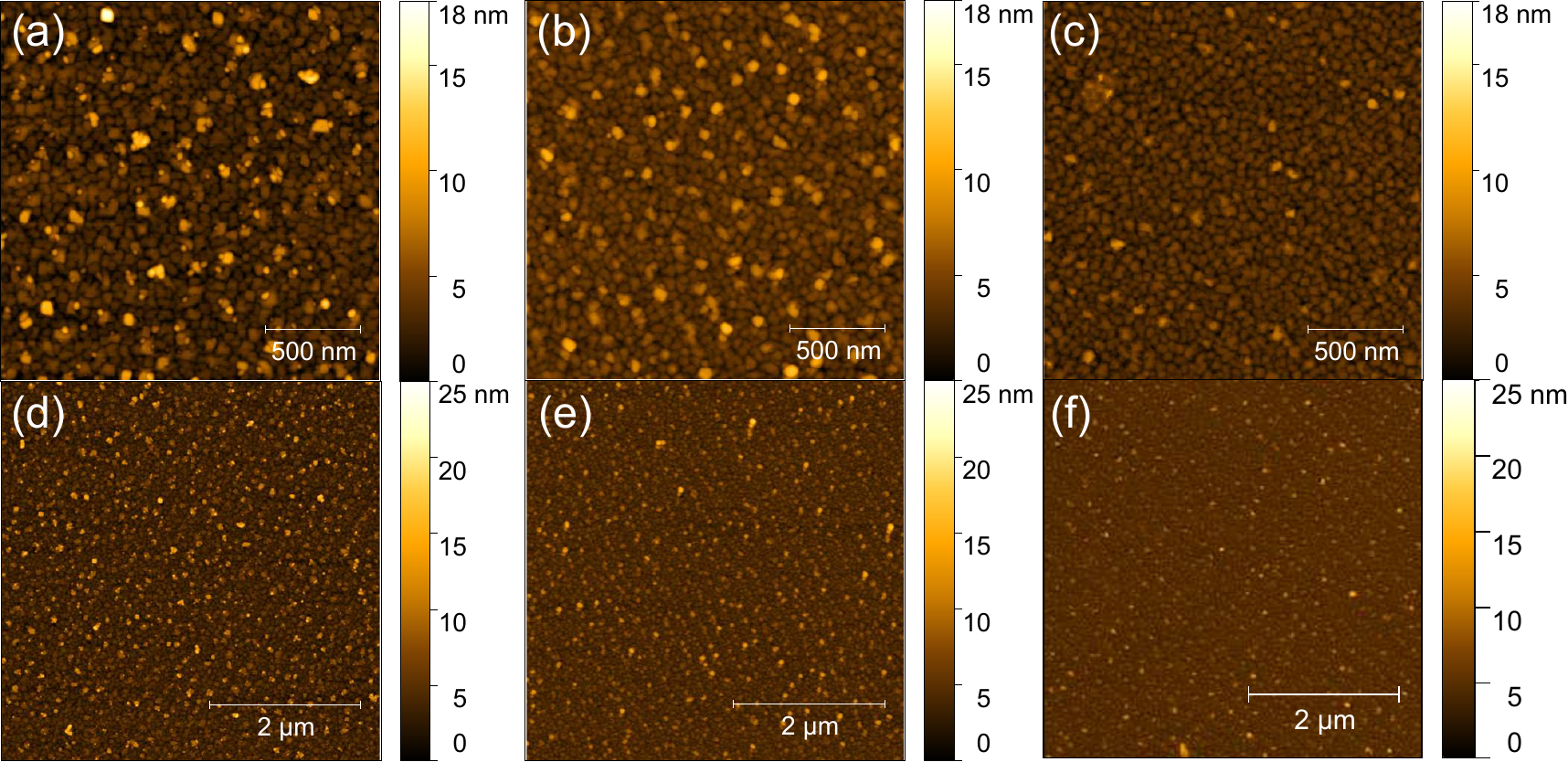}
\caption{\label{fig:SI6} (a),(d) AFM images of the same untreated 22.5 nm thick \ch{KTaO3} film shown in Figs. S1 and S3 grown using a \ch{Ta2O5} source contained within an effusion cell. This film exhibts an rms of 1.8 nm over the whole  ($5\times5$) $\upmu$m$^2$ scan. (b),(e) The same sample after rinsing in cold deionized water for 10 mins. (c),(f) The same sample after 10 min in cold deionized water followed by an additonal 10 min rinse in hot deionized water.}
\end{figure}

The experimental results presented in the main manuscript suggest that under some growth conditions potassium oxide accumulates on the surface of the film. Here we consider whether this is plausible by first considering the vapor pressure of the this phase and the substrate temperature used for the growth of the \ch{KTaO3} films. The vapor pressure of potassium oxide depends on the oxygen partial pressure. The \ch{KTaO3} films are grown using the $\sim$10\% ozone output of a commercial ozone generator, with the 90\% balance being molecular oxygen. To estimate the vapor pressure of potassium oxide, it is necessary to convert the ozone partial pressure used for film growth to an equivalent oxygen partial pressure. The equivalent oxygen partial pressure in the system can be estimated from Fig. 1 of Nair \textit{et al.}\cite{Nair2018} 10 \% ozone was supplied in a 90 \% stream \ch{O2} with a pressure of $10^{-6}$ Torr at 650 °C, meaning that the appropriate line to follow in the figure is the $10^{-7}$ line displayed in the aforementioned article. This equates to an oxygen partial pressure of 300-500 Torr. We were unable to find the vapor pressure of potassium oxide under this exact oxygen partial pressure in the literature, but  Lamoreaux and Hildenbrand calculated the vapor pressure at a very close oxygen partial pressure (150 Torr).
\cite{lamoreaux1984}
\\[12pt]
Under 150 Torr of  \ch{O2}  the stable phase is molten \ch{KO2}. The dominant species in the vapor at the substrate temperature used for film growth, 625 °C, is KO and it is calculated to have a vapor pressure of about $10^{-6}$ Torr. Using the same kinetic theory of gases formula mentioned above with a mass corresponding to KO and a substrate temperature of 625 °C, we get a maximum evaporation flux (corresponding to an evaporation coefficient $\alpha$ = 1) of the KO $\Phi_{leaving} \sim$ 1.6$\times 10^{14}$ atoms/$\text{cm}^2$/s. This is within a factor of two of the incident flux of potassium ($\Phi \sim$(4-7)$\times 10^{13}$ atoms/$\text{cm}^2$/s). As evaporation coefficients are typically significantly lower than one ($\alpha$ = 0.1 is typical), it is conceivable that the flux of potassium being supplied during growth exceeds the flux of KO emanating from potassium oxide. This would lead to the accumulation of potassium oxide, in line with our observations. The key point is that the vapor pressure of species over the K-O system are extremely dependent on the oxygen pressure.
The above analysis leads to us suspect that excess potassium oxide remains on the surface after the growth is finished and that this is the grounds for the rough surface by AFM. In an attempt to clean the surface, the sample is first rinsed in cold deionized water followed by AFM and then  rinsed in hot ($\approx$ 60 °C) deionized water. The results are shown in Fig. \ref{fig:SI6}. The cold water rinse (Figs.~\ref{fig:SI6}(b),(e)) roughens the surface further, increasing the rms to 2.0 nm compared to the previous 1.8 nm for the untreated sample. Atomic steps can be observed. The increased roughness suggests that the potassium oxide reacts with the  cold water, but does not removes it. The result of rinsing in deionized water heated to $\approx$ 60 °C is shown in Figs.~\ref{fig:SI6}(c)  and ~\ref{fig:SI6}(f), where the rms is seen to be reduced to 0.9 nm. The interface is noticeably smoother and atomic steps can be clearly observed. This result is in agreement with observations reported for the preparation of \ch{KTaO3}  single crystal substrates with a step-and-terrace surface morphology.\cite{tomar2018} The film shown in Fig. S1 and S4 had \ch{KO_x} on the surface. As we are exploring the possibility of post-growth potassium desorption from the surface of films, we  deliberately cooled the film with the potassium shutter kept open down to the substrate thermocouple temperature of 300 °C.  
\section{Comparing suboxide and e-beam sources to supply tantalum}
In this section we compare \ch{KTaO3} films grown using suboxide and tantalum e-beam sources. Figure S1 compares the XRD of  \ch{KTaO3} films grown with suboxide and e-beam  sources under otherwise similar conditions (same substrate, substrate temperature, potassium flux, background pressure, and growth rate) on \ch{SrTiO3} (001) substrates. Figures S2 and S3 show the XRD $\theta$-2$\theta$ scans, RSMs, and AFM surface morphology  of a film  grown using an e-beam source for tantalum and a suboxide source for \ch{TaO2} (gas), respectively. The AFM images reveal a smoother  surface for the  e-beam film. The AFM image of the suboxide film reveals potassium oxide  residue on the surface. This could be due to the presence of additional oxygen, leading to a higher sticking coefficient of potassium. The rms roughness of the e-beam film is $\approx$ 0.3 nm compared to $\approx$ 1.8 nm for the suboxide film. Another explanation for the differences in roughness could be the different thicknesses of the two samples. It is important to point out that an equivalent surface roughness can be achieved with the suboxide source in thicker films grown on \ch{GdScO3} (110)$_o$ (See Fig. 1). The \ch{KTaO3} film grown using the  tantalum e-beam source is strained to the \ch{SrTiO3} (001) substrate despite the large lattice mismatch between \ch{KTaO3} and \ch{SrTiO3} ($\approx$ -2.1 \%). 
RSMs around the \ch{SrTiO3} and \ch{KTaO3} 103 reflections  reveal that only the e-beam film is fully strained and the suboxide film is partially relaxed. The underlying reason could be simply due to the difference in film thickness:  22.5 nm vs 10.5 nm for the suboxide and e-beam films, respectively.
\begin{figure}[H]
\centering
\includegraphics[width=0.9\textwidth]{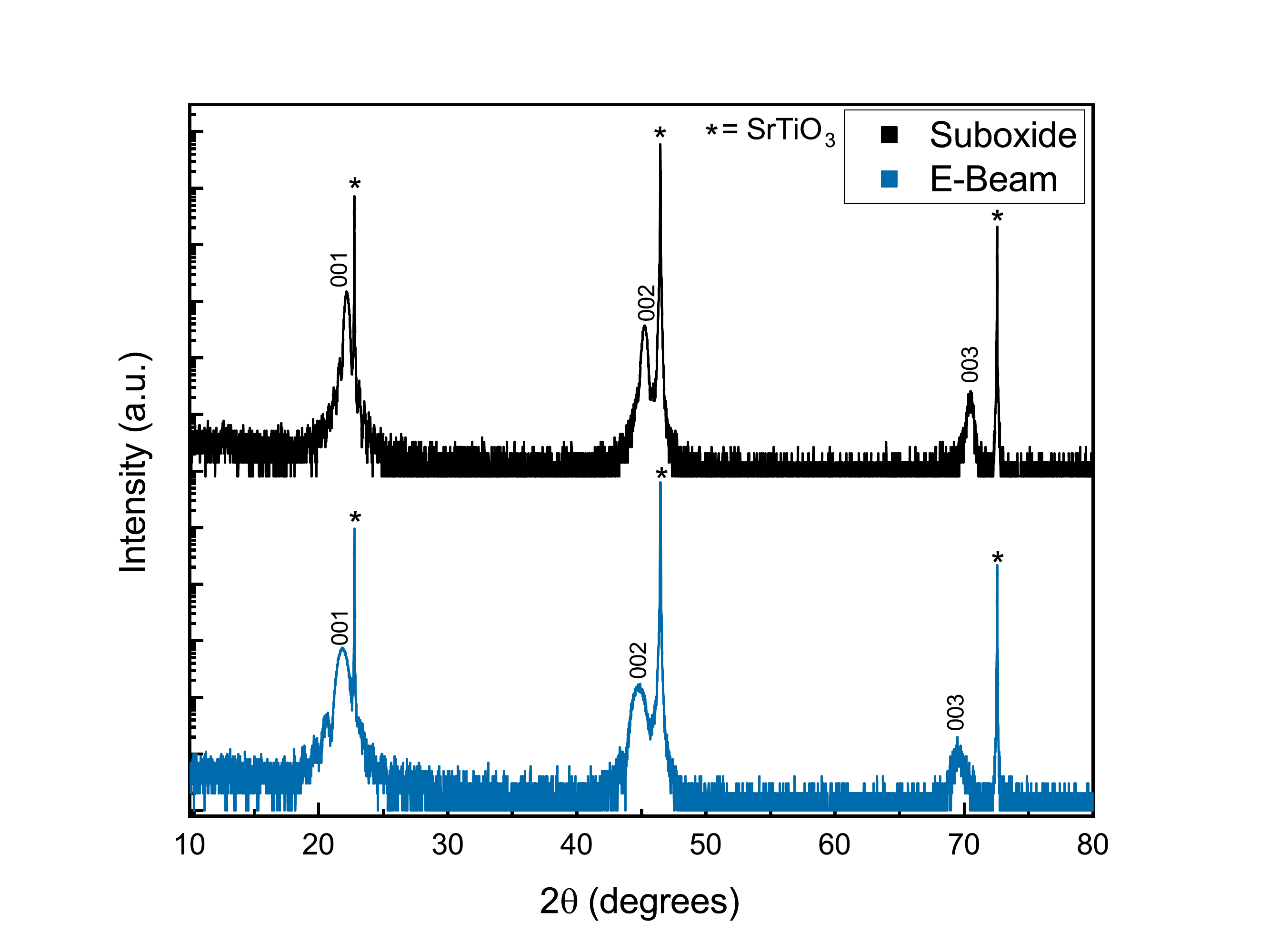}
\caption{\label{fig:SI1} $\theta$-2$\theta$ X-ray diffraction scans of two \ch{KTaO3} films grown on \ch{SrTiO3} (001) substrates, revealing 00$\mathcal{l}$ peaks of \ch{KTaO3}. The \ch{KTaO3} film grown with a suboxide (\ch{TaO2}) molecular beam is 22.5 nm thick. The \ch{KTaO3} film grown using the elemental tantalum molecular beam is 10.5 nm thick. Laue fringes in the vicinity of the 001 \ch{KTaO3} peaks suggest abrupt film-substrate interfaces. Asterisks (*) denote substrate reflections. }
\end{figure}
\begin{figure}[H]
\centering
\includegraphics[width=\textwidth]{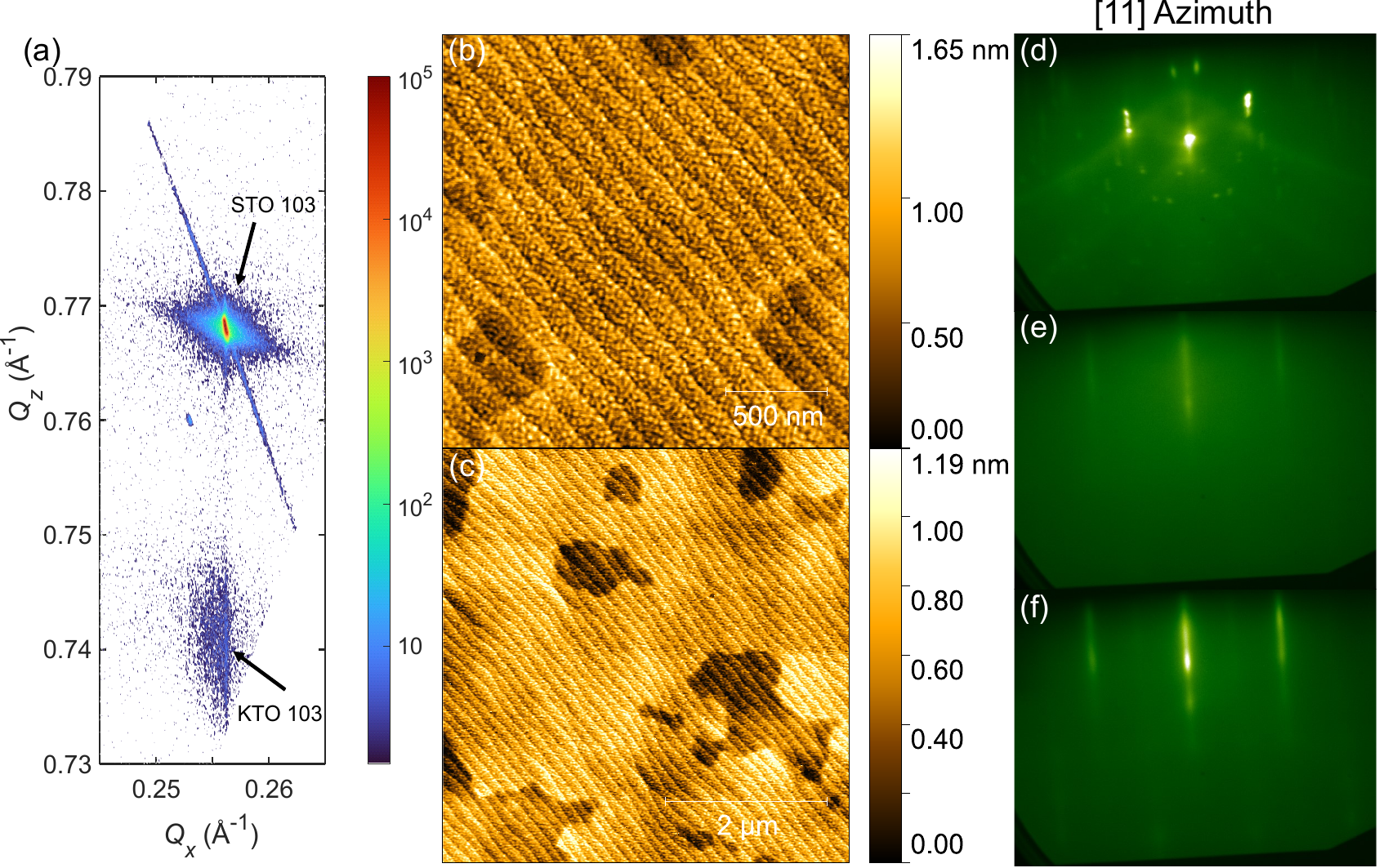}
\caption{\label{fig:SI2} The same 10.5 nm thick \ch{KTaO3} film shown in Fig. S1 grown using a tantalum e-beam source. (a) RSM of the \ch{SrTiO3} and \ch{KTaO3} 103 peaks. (b),(c) AFM images at different magnifications. The root mean square (rms) over the ($5\times5$) $\upmu$m$^2$ scan is 0.26 nm. (d) RHEED patterns of the bare substrate, (e) at the beginning and (f) the end of the growth of the 10.5 nm thick \ch{KTaO3} film viewed along the [11] azimuth.}
\end{figure}
\begin{figure}[H]
\centering
\includegraphics[width=\textwidth]{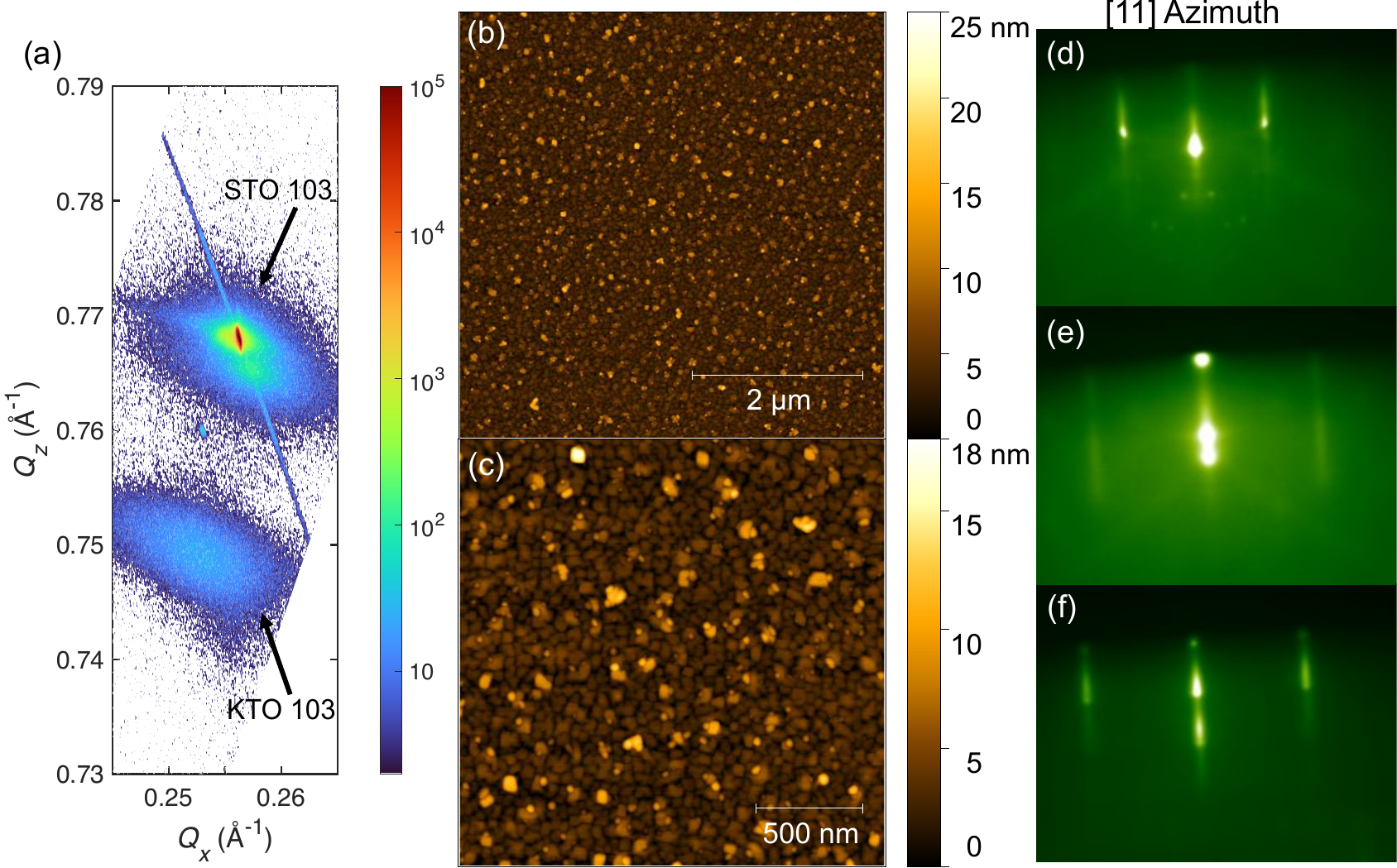}
\caption{\label{fig:SI3} The same 22.5 nm thick \ch{KTaO3} film shown in Fig. S1 grown using  a \ch{Ta2O5} source contained witin an effusion cell. (a) RSM of the \ch{SrTiO3} and \ch{KTaO3} 103 peaks. (b),(c) AFM images at different magnifications. The root mean square (rms) over the ($5\times5$) $\upmu$m$^2$ scan is 1.8 nm. (d) RHEED patterns of the bare substrate, (e) at the beginning and (f) the end of the growth of the 22.5 nm thick \ch{KTaO3} film viewed along the [11] azimuth.}
\end{figure}
\section{Confirmation of epitaxy and lack of twinning}
\begin{figure}[h]
\centering
\includegraphics[width=\textwidth]{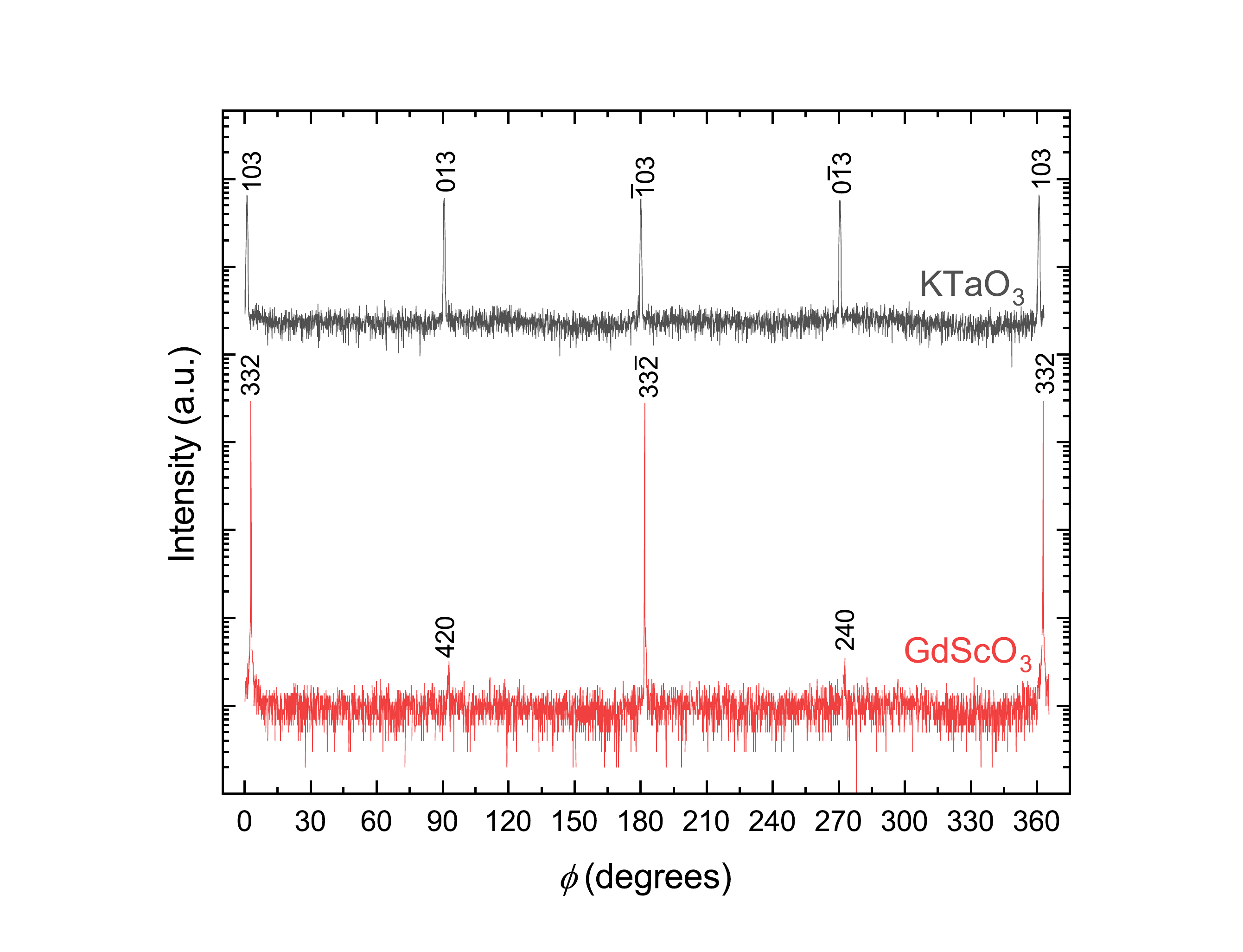}
\caption{\label{fig:SI4} 103 $\phi$-scan of the 18 nm thick \ch{KTaO3} film grown on a \ch{GdScO3} (110)$_o$ substrate
with an effusion cell containing \ch{Ta2O5} that was shown in the main article in Figs. 1, 2, and 3 as well as in Figs.S7, S8, and S9. Peaks of the \ch{GdScO3} substrate designated as 420 and 240 are shoulders of the actual reflections.}
\end{figure}

For the film shown in Figs. 1-3 of the main article, a $\phi$-scan was done to confirm epitaxy and investigate the possibility of twinning. The results in Fig.~\ref{fig:SI4} show the expected 4-fold symmetry of a single-domain epitaxial \ch{KTaO3} film.

\section{Confirmation that the \ch{KTaO3} films are not contaminated with indium}
\begin{figure}[H]
\centering
\includegraphics[width=\textwidth]{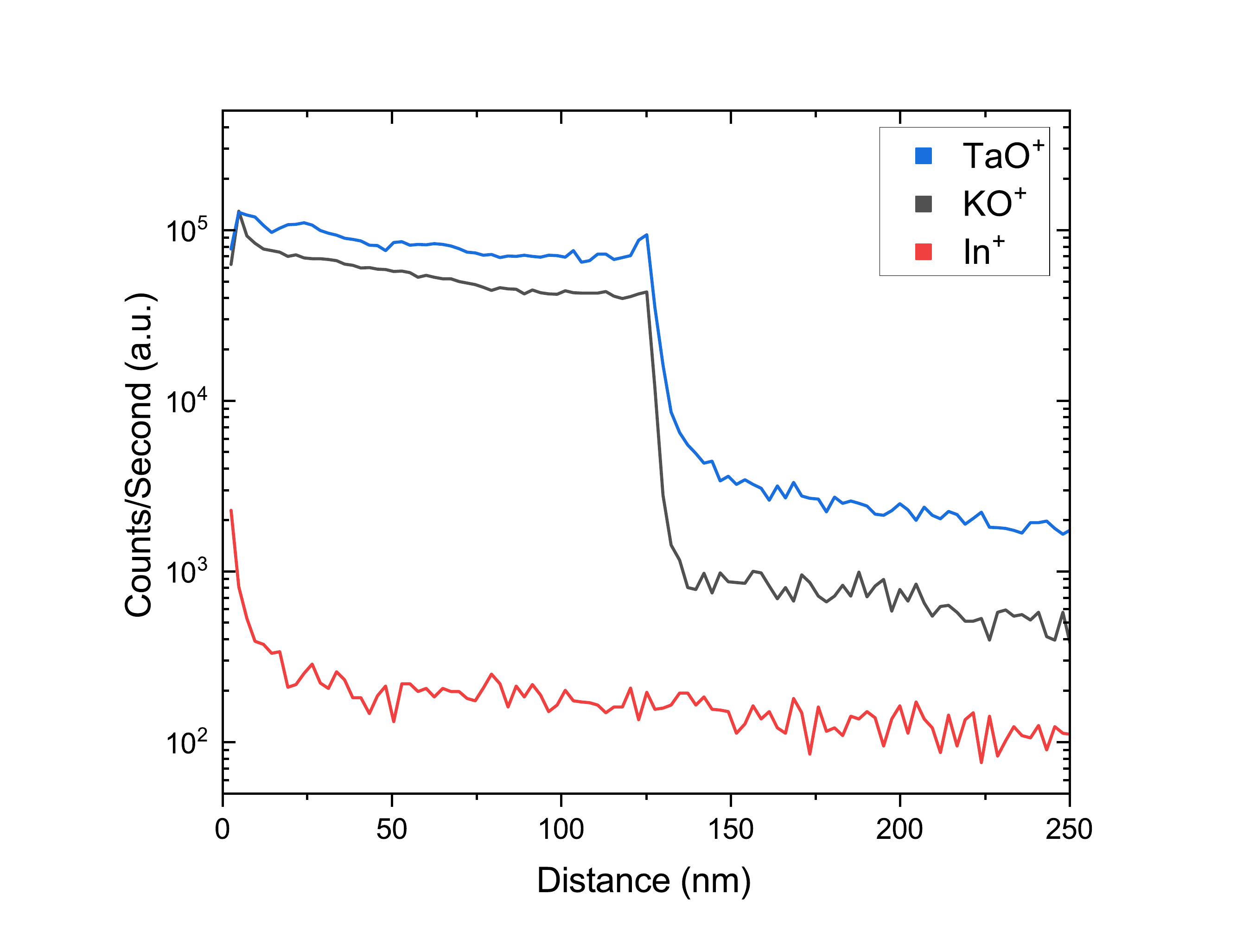}
\caption{\label{fig:SI5} Secondary ion mass spectrometry (SIMS) depth profile into a 130 nm thick \ch{KTaO3} film grown using \ch{InK4} and \ch{Ta2O5} sources on a \ch{SrTiO3} (001) substrate. 
The depth profile includes the entire \ch{KTaO3} film  and well beyond into the \ch{SrTiO3} (001) substrate. No indium is seen by SIMS.}
\end{figure}

The \ch{KTaO3} films were grown using a potassium molecular beam emanating from a \ch{KIn4} intermetallic source. Although our expectation based on vapor pressures was that the resulting molecular beam should be pure potassium, this was explicitly tested by analyzing a \ch{KTaO3} film grown using \ch{KIn4} and \ch{Ta2O5} sources. Secondary ion mass spectrometry (SIMS) of a 130 nm thick \ch{KTaO3} film on grown on a  \ch{SrTiO3} (001) substrate was measured. The sample was depth profiled using the dynamic SIMS mode with an oxygen \ch{O2^+} primary beam. The multiplicative conversion factor (SIMS sputter rate) is 5.105 $\times 10^{-2}$ nm/s. Figure~\ref{fig:SI5} shows the results, from which it can be seen that the indium signal stays at the same level well into the substrate material (\ch{SrTiO3}).
SIMS is known to detect elemental species down to a detection limit of ppb for typical semiconductors.\cite{stevie2015secondary} The count rate for the more sensitive \ch{In^+} species, compared to \ch{InO^+}, monitored in this oxide semiconductor is orders of magnitude lower than those for \ch{KO^+} amd \ch{TaO^+} in the \ch{KTaO3} matrix. This indicates that the indium contamination is low. For an \ch{O2^+} primary beam, EAG reports\cite{eag} a sensitivity of 5 $\times 10^{14}$ atoms/$\text{cm}^3$ for \ch{B} and \ch{Sn} in \ch{Ga2O3}. \ch{B} has the same valence and \ch{Sn} is close in atomic mass to \ch{In}, respectively. Without a standard, the impurity level cannot be quantified, but it is likely in the ppb range given the EAG data for \ch{Ga2O3}.

\section{STEM-EDX of the film-substrate interface}
\begin{figure}[H]
\centering
\includegraphics[width=\textwidth]{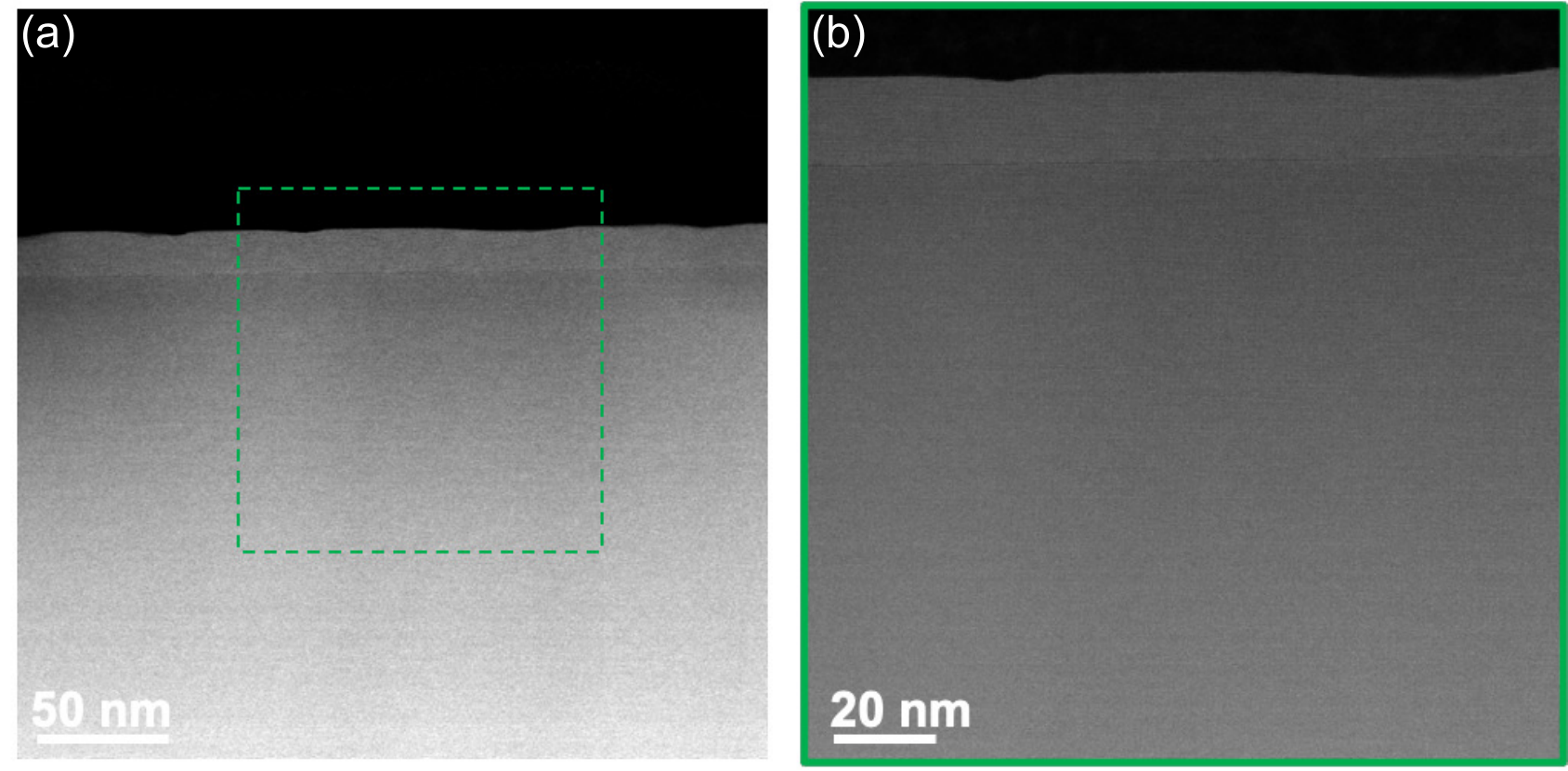}
\caption{\label{fig:SI7} Low-magnification HAADF-STEM of the same 22.5 nm thick \ch{KTaO3} film on \ch{GdScO3} (110)$_o$ shown in Figs. S1, S3, and S4, grown using a \ch{Ta2O5} source contained within an effusion cell. No extended defects, e.g., threading dislocations, are seen.}
\end{figure}
\begin{figure}[h]
\includegraphics[width=\textwidth]{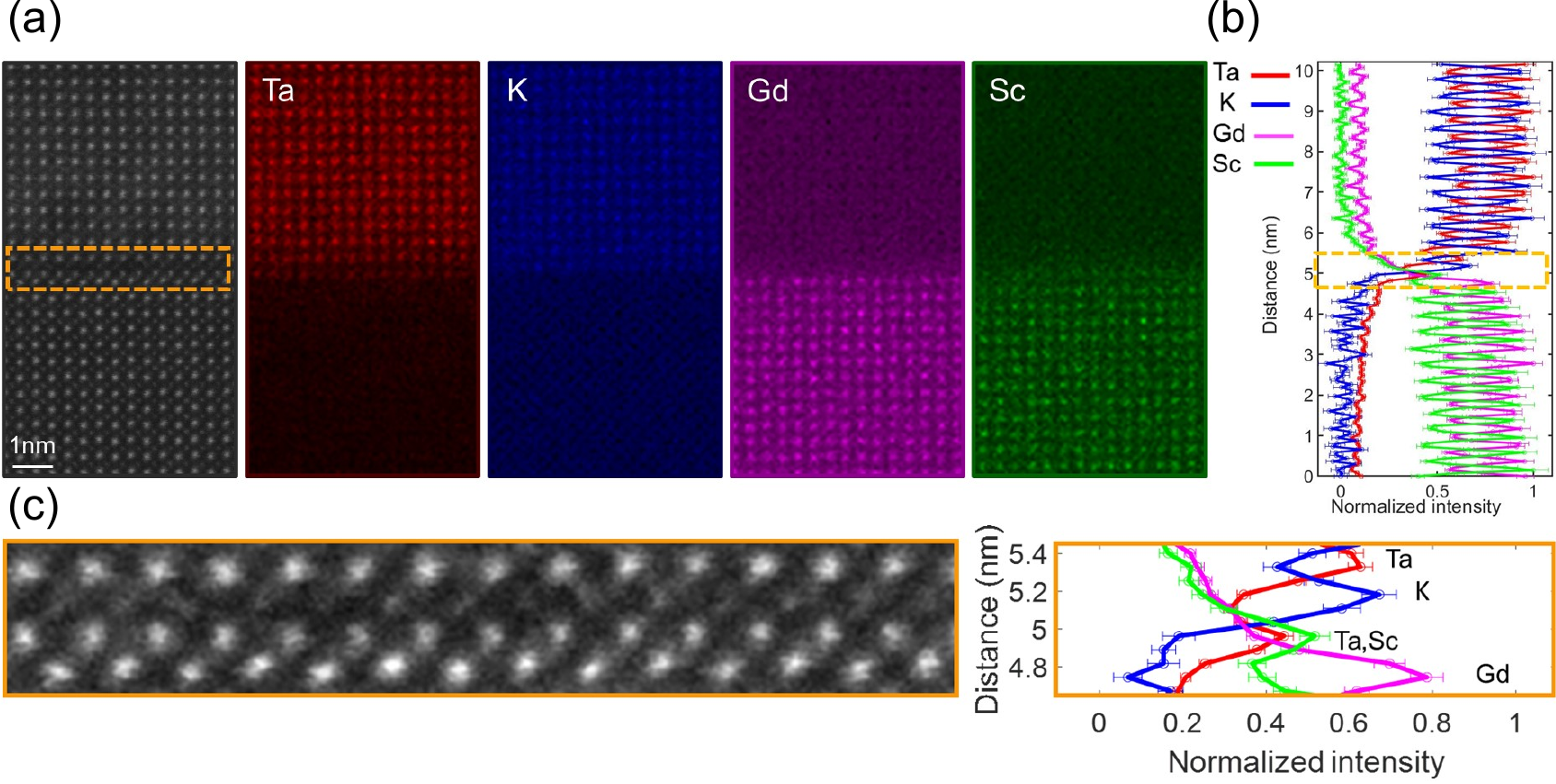}
\caption{\label{fig:SI8} (a) HAADF image and EDX elemental map of potassium, tantalum, gadolinium, and scandium across the interface of \ch{KTaO3} and \ch{GdScO3} grown using a \ch{Ta2O5} source contained within an effusion cell. (b) Line profile of the normalized concentrations of the K, Ta, Gd, and Sc along the interface. (c) Zoomed-in interface and line profile showing the formation of \ch{Ta_{$y$}Sc_{$1-y$}O2} ($y\approx$ 0.46) and KO at the interface. }
\end{figure}
\begin{figure}[H]
\centering
\includegraphics[width=0.5\textwidth]{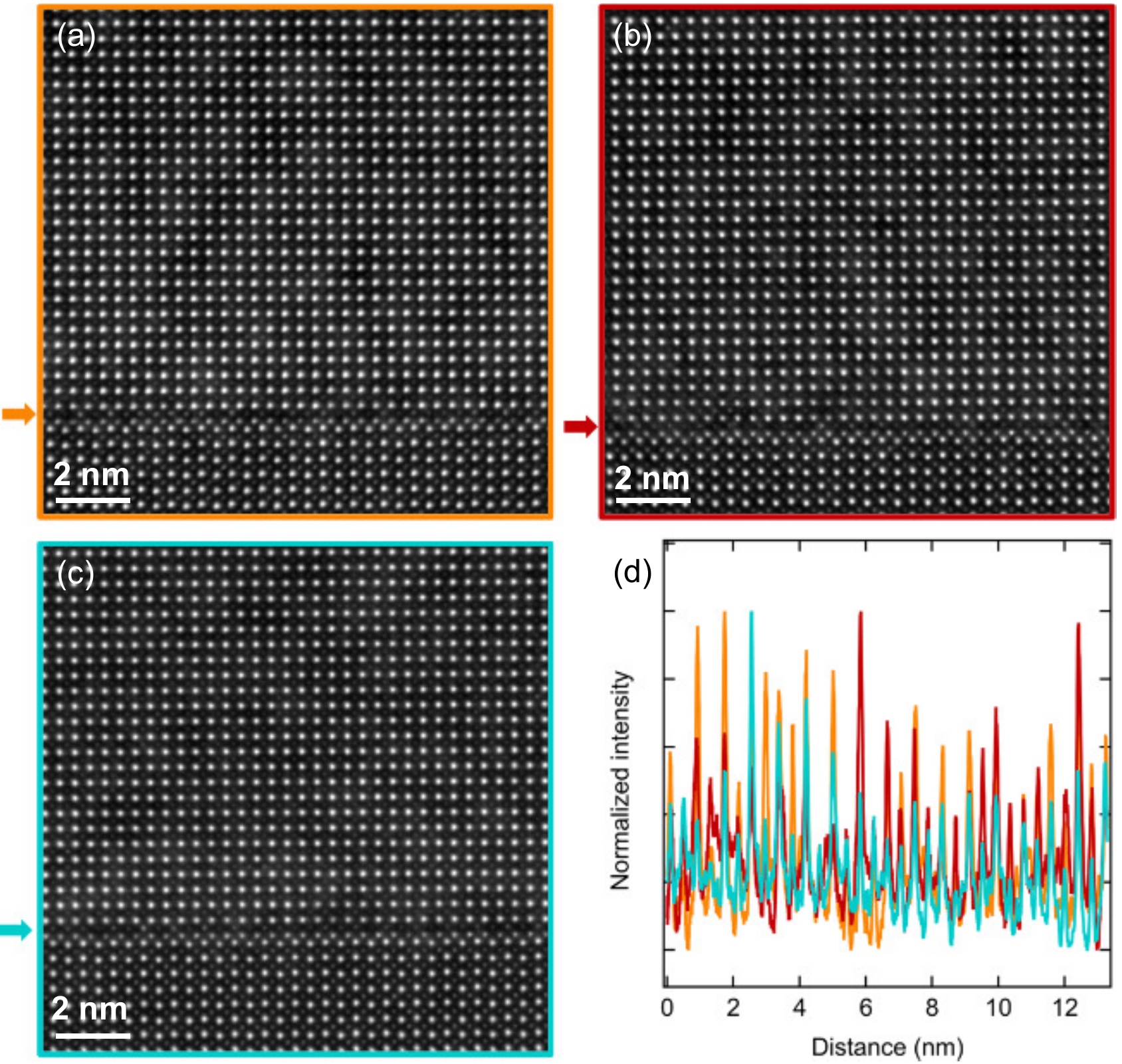}
\caption{\label{fig:SI9} (a)-(c) HAADF-STEM images of different regions of the  film grown on \ch{GdScO3} (110)$_o$ grown with  a \ch{Ta2O5} source contained within an effusion cell with different interface structures. (d) Line profiles obtained from the second layer of the interface (indicated by the positions of the arrow of the correspond colors) from (a)-(c), show occasional brighter contrast at the KO layer of the interface.\cite{knuthwebsite} }
\end{figure}

 An EDX elemental map is used to investigate the interface structure of the same \ch{KTaO3}/\ch{GdScO3} film shown in Fig. 3. Figure S5(a)  shows an HAADF-STEM image of a selected region at the \ch{KTaO3}/\ch{GdScO3} interface and a corresponding high-resolution EDX elemental map. The EDX map reaveals an intermixed region at the interface. Figure~\ref{fig:SI8}(b) shows averaged line scans across the interface for the tantalum, potassium, gadolinium, and scandium elements. The zoomed-in EDX line profile and corresponding HAADF-STEM are shown in Fig.~\ref{fig:SI8}(c).  The scandium and tantalum line profiles overlap at the first layer of the interface, forming Ta$_{y}$Sc$_{1-y}$O$_{2}$. EDX quantification suggests $y$ $\approx$ 0.46, which makes the first layer charge neutral. The EDX profile shows that the second layer of the interface predominantly consists of potassium atoms forming \ch{K^{1+}O^{2-}}. The formation of KO will yield a net charge of (-1) at the interface and will conserve the polarity of the system.
HAADF-STEM images of the other regions of the sample (Figs.~\ref{fig:SI9}(a)-(c)) show that the top layer of the interface does not always have uniform intensity. Figure~\ref{fig:SI9}(d) shows the intensity line profile of the second interface layer obtained from the regions highlighted by an arrow in Figs.~\ref{fig:SI9}(a)-(c). It is clear from the intensity profile that the marked layers have occasional higher intensity columns as well.  The brighter contrast in these layers can be related to gadolinium atoms in K$_{x}$Gd$_{1-x}$O. Formation of the K$_{x}$Gd$_{1-x}$O layer with $x$  $>$ 0.5 will lead to a net charge approaching (-1) at the interface as $x$ approaches 1. 

\bibliography{sup.bib}